\documentclass[twocolumn,aps,superscriptaddress]{revtex4}
\usepackage{graphicx}
\usepackage{float}
\usepackage{dcolumn}
\usepackage{bm}
\usepackage{color}
\usepackage{amsmath}
\usepackage{amssymb}
\usepackage{amsfonts}
\usepackage{esint}
\usepackage{times}
\usepackage{xcolor}
\usepackage{phaistos}
\usepackage{braket}
\usepackage{comment}

\usepackage{pdfpages}

\usepackage[colorlinks,linkcolor=blue,anchorcolor=blue,citecolor=blue]{hyperref}

\begin{document}

\title{Baseband control of superconducting qubits with shared microwave drives}

\author{Peng Zhao}
\email{shangniguo@sina.com}
\affiliation{Beijing Academy of Quantum Information Sciences, Beijing 100193, China}
\author{Ruixia Wang}
\email{wangrx@baqis.ac.cn}
\affiliation{Beijing Academy of Quantum Information Sciences, Beijing 100193, China}
\author{Meng-Jun Hu}
\affiliation{Beijing Academy of Quantum Information Sciences, Beijing 100193, China}
\author{Teng Ma}
\affiliation{Beijing Academy of Quantum Information Sciences, Beijing 100193, China}
\author{Peng Xu}
\affiliation{Institute of Quantum Information and Technology,
Nanjing University of Posts and Telecommunications, Nanjing, Jiangsu 210003, China}
\author{Yirong Jin}
\affiliation{Beijing Academy of Quantum Information Sciences, Beijing 100193, China}
\author{Haifeng Yu}
\email{hfyu@baqis.ac.cn}
\affiliation{Beijing Academy of Quantum Information Sciences, Beijing 100193, China}

\date{\today}

\begin{abstract}
Accurate control of qubits is the central requirement for building functional quantum processors. For the current superconducting quantum processor, high-fidelity control of qubits is mainly based on independently calibrated microwave pulses, which could differ from each other in frequencies, amplitudes, and phases. With this control strategy, the needed physical resource could be challenging, especially when scaling up to large-scale quantum processors is considered. Inspired by Kane's proposal for spin-based quantum computing, here, we explore theoretically the possibility of baseband flux control of superconducting qubits with only shared and always-on microwave drives. In our strategy, qubits are by default far detuned from the drive during system idle periods, qubit readout and baseband flux-controlled two-qubit gates can thus be realized with minimal impacts from the always-on drive. By contrast, during working periods, qubits are tuned on resonance with the drive and single-qubit gates can be realized. Therefore, universal qubit control can be achieved with only baseband flux pulses and always-on shared microwave drives. We apply this strategy to the qubit architecture where tunable qubits are coupled via a tunable coupler, and the analysis shows that high-fidelity qubit control is possible. Besides, the baseband control strategy needs fewer physical resources, such as control electronics and cooling power in cryogenic systems, than that of microwave control. More importantly, the flexibility of baseband flux control could be employed for addressing the non-uniformity issue of superconducting qubits, potentially allowing the realization of multiplexing and cross-bar technologies and thus controlling large numbers of qubits with fewer control lines. We thus expect that baseband control with shared microwave drives can help build large-scale superconducting quantum processors.

\end{abstract}

\maketitle


\section{Introduction}\label{SecI}

For quantum processors built with superconducting qubits, both the control accuracy and the qubit
number have shown steady improvement over the past two decades \cite{Kjaergaard2020}.
Notably, quantum gate operations, which are generally implemented by using microwave
or baseband flux pulses \cite{Krantz2019}, with errors reaching
the fault-tolerant thresholds have been achieved in quantum
processors with several tens of qubits \cite{Arute2019,Zhu2022,Acharya2022,Kim2021}.
Nevertheless, it is known that fulfilling the full promises of quantum computing requires
the implementation of fault-tolerant schemes, which will need the high-fidelity control
of millions of qubits \cite{Fowler2012,Gidney2021}. In such large-scale superconducting
quantum processors, the needed physical resource, such as control electronics and cooling power in
cryogenic systems, could be the most challenging obstacle for achieving accurate control
of qubits, let alone solving the wiring problem \cite{Frankea2019,Reilly2019,Martinis2020}
and the device yield problem \cite{Hertzberg2021,Kreikebaum2020}.

In superconducting quantum processors each qubit has typically a
different set of parameters: frequency, anharmonicity, coupling
efficiency and signal attenuations in control lines.
Thus, in current small-scale quantum processors, to ensure accurate qubit control, each
qubit should have its dedicated control pulse with different parameter
settings \cite{Kelly2018,Klimov2020}. This means that the microwave control pulses
could differ from each other in their amplitudes, frequencies, and phases, while for
the baseband flux pulse, their amplitudes could be different. Moreover, these control pulses
are generated at room temperature, and then delivered to qubits in the cryogenic system
through a series of attenuators and filters for suppressing harmful noises, such as
thermal noise \cite{Krantz2019,Chen2018,Krinner2019}. Considering these general arguments, the
physical resource for realizing qubit control could be highly related to the type of
employed control. To be more specific, the microwave control and its signal synthesis
are more complicated and expensive than that of the baseband flux control, for which only a
single digital-to-analog converter (DAC) per qubit is needed \cite{Krantz2019,Chen2018,Krinner2019}.
Moreover, given the limited available cooling power in cryogenic systems, the microwave
control lines generally need the attenuation of $60\,\rm dB$ (about $20\,\rm dB$ at
the mixing chamber plate (MXC), for which the available cooling power is smallest),
leading to heating loads larger than that of the baseband flux lines (about $20\,\rm dB$,
need no attenuation at the MXC stage) \cite{Arute2019,Chen2018,Krinner2019}. Additionally, in
large-scale quantum processors, microwave control requires higher-density control lines, making it
challenging to suppress the microwave crosstalk \cite{Wenner2011,Rosenberg2019,Huang2021M} and
thus to achieve high-fidelity qubit control. By contrast, with baseband flux control, there
exists only a single control parameter, potentially allowing the application of multiplexing
technologies and cross-bar technologies to address the challenges, e.g., the wiring problem, toward
large-scale quantum computing \cite{Hill2015,Vandersypen2017,Veldhorst2017,Li2018}.

\begin{figure}[tbp]
\begin{center}
\includegraphics[keepaspectratio=true,width=\columnwidth]{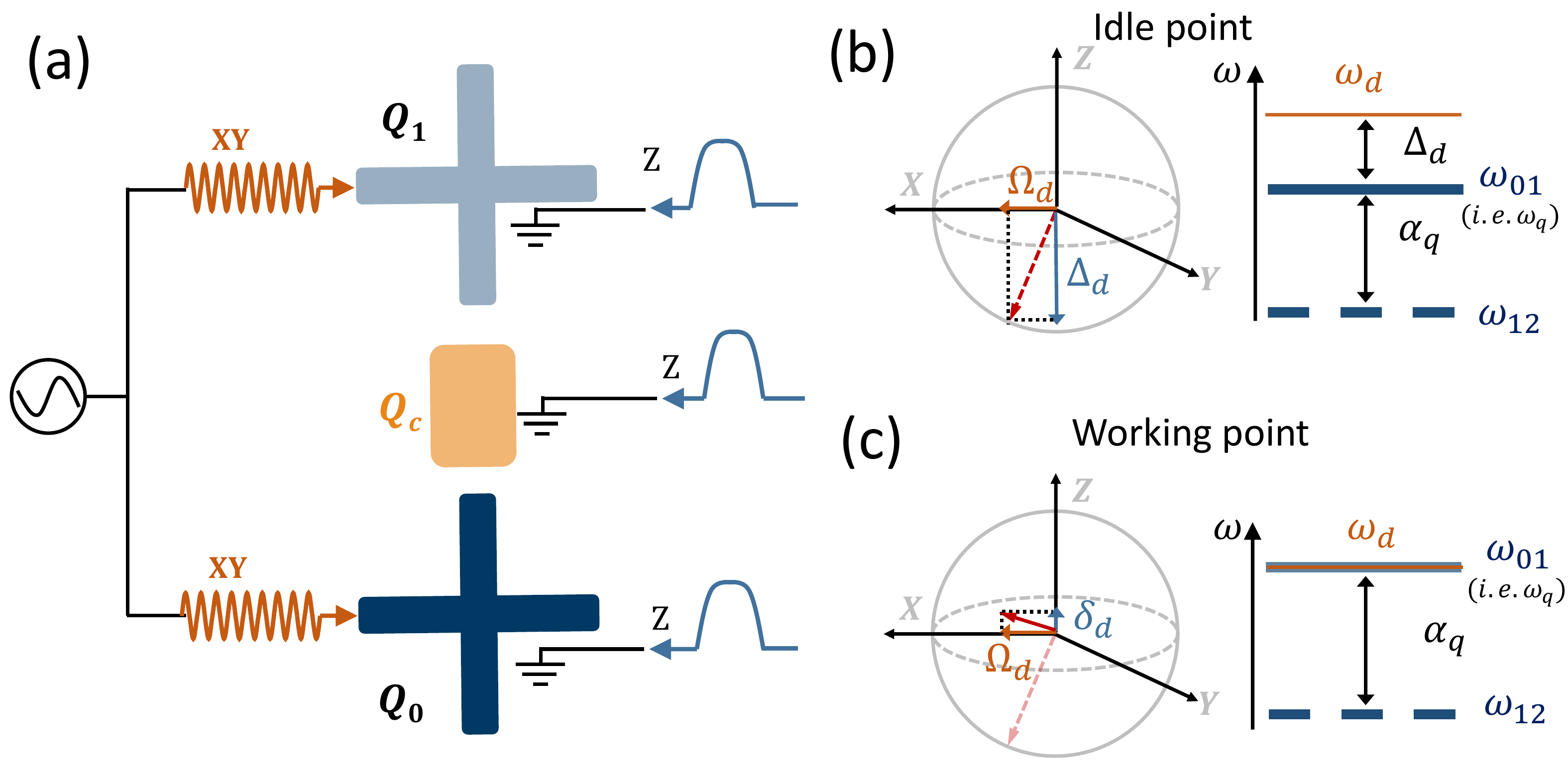}
\end{center}
\caption{Baseband flux control of transmon qubits with a shared always-on microwave drive.
(a) Sketch of a baseband flux controlled two frequency-tunable transmon qubits ($Q_{0}$ and $Q_{1}$) with
dedicated Z lines. The two qubits are coupled via a coupler $Q_{c}$, which could be a tunable coupler.
Through a shared XY line, the two qubits are driven simultaneously by a global and always-on microwave
drive with the frequency $\omega_{d}$ and the constant amplitude $\Omega_{d}$. Baseband flux
pulses are delivered to the qubits and coupler through their dedicated Z lines. (b) At the idle point, the
qubit is far detuned from the drive, i.e., the drive
detuning, $|\Delta_{d}|=|\omega_{01}-\omega_{d}|\gg\Omega_{d}$. Left: Bloch vector in the rotating frame with respect to
the drive (here, confined to qubit subspace spanned by the lowest two-energy levels of
the transmon qubit). Due to the always-on drive, the Bloch vector (dashed red arrow) at the idle point is tilted toward the
X-axis. We thus choose the logical computational states to be the dressed eigenstates defined by
the tilted Bloch vector. Right: Energy level diagram of
the qubit at the idle point. Here, $\alpha_{q}$ denotes the qubit anharmonicity. (c) When
operating the system at the working point, where the qubit is on-resonance with the drive, single-qubit
operations can be implemented. Left: Bloch vector (solid red arrow) at the working point. Since the initial Bloch vector is
slightly tilted, a small detuning $\delta_{d}$ between the qubit and the drive is needed for enabling
complete Rabi oscillations. Right: Energy level diagram of the qubit at the working point.}
\label{fig1}
\end{figure}

Given the above discussion, when scaling up to large-scale quantum processors, implementing
baseband flux control could make requirements less stringent than that of microwave control.
However, currently, microwave control is generally the essential one for implementing qubit
addressing and single-qubit gate operations \cite{Motzoi2009,Chen2016,Krantz2019}, and even
for two-qubit gates, such as cross-resonance gates \cite{Chow2011}. In this work, we explore
theoretically the possibility of developing the baseband flux control of frequency-tunable
qubits with the help of always-on shared microwave drives. It should be noted that previous
works on studying baseband control of superconducting qubits mainly focus on low-frequency
qubits, e.g., the composite qubit \cite{Campbell2020} and heavy-fluxonium qubit \cite{Zhang2021}, here
we focus on the transmon qubits \cite{Koch2007} that have been widely used in current
superconducting quantum processors. The basic idea of our control strategy is sketched
in Fig.~\ref{fig1}(a), where two qubits are coupled via a coupler and the always-on microwave
drive (XY line) is shared by both qubits (in principle, can be extended to
multi-qubit cases), the qubit control and the single-qubit addressing can be realized only
through the flux (Z) control lines. Our work is motivated by Kane's proposal for realizing
spin-based quantum computing \cite{Kane1998}, where the spin qubit is by default
off-resonance with the global always-on microwave magnetic
field (i.e., at the idle point) and single-qubit gate operations are realized by tuning
the spin qubit on-resonance with the field (i.e., at the working point) \cite{Wolfowicz2014,Laucht2015},
as shown in Figs.~\ref{fig1}(b) and~\ref{fig1}(c). Due to the always-on
shared drive, the computational states are the basis states of the microwave-dressed
qubit \cite{Liu2006,Zhao2022,Wei2022}, and accordingly, in the present work, all the qubit
control are analyzed based on this microwave-dressed basis. As an example application
of this baseband control strategy, in a system comprising two frequency-tunable transmon
qubits coupled via a tunable coupler \cite{Yan2018}, we study the feasibility of this
strategy for achieving high-fidelity gate operations. By theoretical analysis, we will
show that:

(i) To implement single-qubit gate operations, especially, $\sqrt{X}$ gates, the
baseband Z(flux)-control provides great flexibility in the gate tune-up procedure. This
flexibility could be used to relieve stringent requirements on qubit frequency, drive strength,
and gate time for implementing single-qubit gates, and thus can even compensate for the
non-uniformity of qubit parameters, potentially allowing to perform multiplexed
control of qubits.

(ii) Since the transmon qubit has a weak anharmonicity, in the traditional microwave
control setup, leakage during gate operations can be suppressed by using the
derivative removal by adiabatic gate (DRAG) scheme \cite{Motzoi2009}. In our setup, while the DRAG scheme
can no longer be directly utilized, we show that by using a modified fast-adiabatic
scheme, the leakage can also be suppressed heavily.

(iii) While the always-on microwave drive is detuned from the qubits, it can
induce ac-Stark frequency shifts on the qubits \cite{Tuorila2010,Schneider2018,Liu2006,Zhao2022,Wei2022}.
Consequently, any fluctuations in the drive amplitude will cause qubit dephasing. By
numerical simulation, we study the effect of the amplitude-dependent noise on
the qubit and show that the fluctuation-induced dephasing can be
eliminated by tuning the qubit away from the drive. Similarly, by numerical simulation
of qubit readout dynamics, the impacts of the always-on drive on the
readout fidelity can also be neglected safely when the drive detuning is far larger
than the drive strength.

(iv) In the qubit architecture with tunable coupling, we show that with the baseband
control strategy and the modified fast-adiabatic scheme, high-fidelity single-qubit
gates are achievable. We also outline the leading error mechanisms that should be
considered carefully when applying baseband control in large-scale quantum
systems. Additionally, we further show that with the always-on microwave
drive, baseband-controlled two-qubit CZ gates can still be achieved with
high gate fidelity and short gate length.

The rest of the paper is organized as follows. In Sec.~\ref{SecII}, we provide an overview of the baseband
control scheme. In Sec.~\ref{SecIII}, we consider an example application of the
baseband control strategy for achieving high-fidelity single- and two-qubit gates in a
qubit architecture with tunable coupling. In Sec.~\ref{SecIV}, we will provide discussions of the challenges
and opportunities for realizing the baseband control strategy in superconducting quantum processors. Finally, we
provide a summary of our work in Sec.~\ref{SecV}.

\section{Overview of the baseband control setup}\label{SecII}

Here, we provide an overview of the baseband control setup schematically illustrated
in Fig.~\ref{fig1}(a). In our setup, frequency-tunable transmon qubits are driven simultaneously by a
single always-on global drive with a constant amplitude and each qubit has its dedicated
flux control lines, i.e., Z lines. Same to Kane's proposal \cite{Kane1998}, single-qubit
addressing or single-qubit gate operations can be implemented by tuning the qubits
on-resonance with the global drive, as shown in Fig.~\ref{fig1}(c). By contrast, when
biasing at the idle point, as shown in Fig.~\ref{fig1}(b), the qubit is far detuned from the
drive, thus in principle, qubit readout and baseband-controlled two-qubit gates can be
realized with minimal impacts from the always-on drive. To evaluate the feasibility of the
control scheme, in the following, we first consider implementing universal control
of an ideal two-level system, which is subjected to an always-on drive, using only Z-control.
Next, we consider a more practical case of transmon qubits, which has a weak
qubit anharmonicity, making qubits particularly susceptible to leakage during gate
operations. We will show that with a fast-adiabatic scheme \cite{Martinis2014b}, Z-controlled single-qubit gate
operations can be achieved with fast speed and low leakage. Finally, by biasing the qubit at the
idle point, we further study the impact of the always-on drive on the qubit dephasing
and qubit readout.

\subsection{Z-control of an ideal two-level system}\label{SecIIA}

For a baseband controlled two-level system (TLS) subjected to an always-on global drive, the system Hamiltonian
is (hereinafter, we set $\hbar=1$)
\begin{equation}
\begin{aligned}\label{eq1}
H_{lab}=\frac{\omega_{q}}{2}\sigma_{z}+\Omega_{d}\cos(\omega_{d}t)\sigma_{x}
\end{aligned}
\end{equation}
where $\omega_{q}$ is the bare qubit frequency and can change according to the Z
control pulse, $\omega_{d}$ and $\Omega_{d}$ are the frequency and
the amplitude of the drive, respectively. Moving into the rotating frame with respect to the
global drive and after applying the rotating wave approximation (RWA), the Hamiltonian
reads
\begin{equation}
\begin{aligned}\label{eq2}
H_{rot}=\frac{\Delta_{d}}{2}\sigma_{z}+\frac{\Omega_{d}}{2}\sigma_{x}
\end{aligned}
\end{equation}
where $\Delta_{d}=\omega_{q}-\omega_{d}$ denotes the drive detuning. Note that
unless otherwise stated, the RWA is used throughout this work.

At the idle point, the drive detuning is far large than the drive strength, thus
the Bloch vector is slightly tilted towards the X-axis, as shown in Fig.~\ref{fig1}(b).
Generally, the dressed eigenstates defined by this tilted Bloch vector are chosen to be
the logical computational states. The tilted angle and the dressed states can be quantitatively
obtained by diagonalization of the Hamiltonian in Eq.~(\ref{eq2}), giving rise to
\begin{equation}
\begin{aligned}\label{eq3}
H_{diag}=\frac{\Delta}{2}Z,\,{\rm with}\,Z\equiv\cos{\theta}\sigma_{z}+\sin{\theta}\sigma_{x},
\end{aligned}
\end{equation}
where $\theta=\arctan(\Omega_{d}/\Delta_{d})$ is the tilted angle and $\Delta=\sqrt{\Delta_{d}^{2}+\Omega_{d}^{2}}$.
From the above discussions, when $|\Delta_{d}|\gg \Omega_{d}$, one can neglect the angle, as well as the
difference between the bare states and the dressed states.

As shown in Eq.~(\ref{eq3}), by biasing the qubit at the idle point, the Z rotations
can be easily realized by choosing suitable delay times $\tau$ between Z pulses, i.e.,
\begin{equation}
\begin{aligned}\label{eq4}
U_{z}=e^{-i\frac{\Delta \tau}{2}Z}.
\end{aligned}
\end{equation}
Note here that compared with the traditional microwave control, Virtual-Z (VZ) gate scheme \cite{McKay2017} is not suitable for the
present baseband control. However, similar to the VZ gate, besides time delay, here, no actual control
pulses are needed for implementing Z rotations.

\begin{figure}[tbp]
\begin{center}
\includegraphics[keepaspectratio=true,width=\columnwidth]{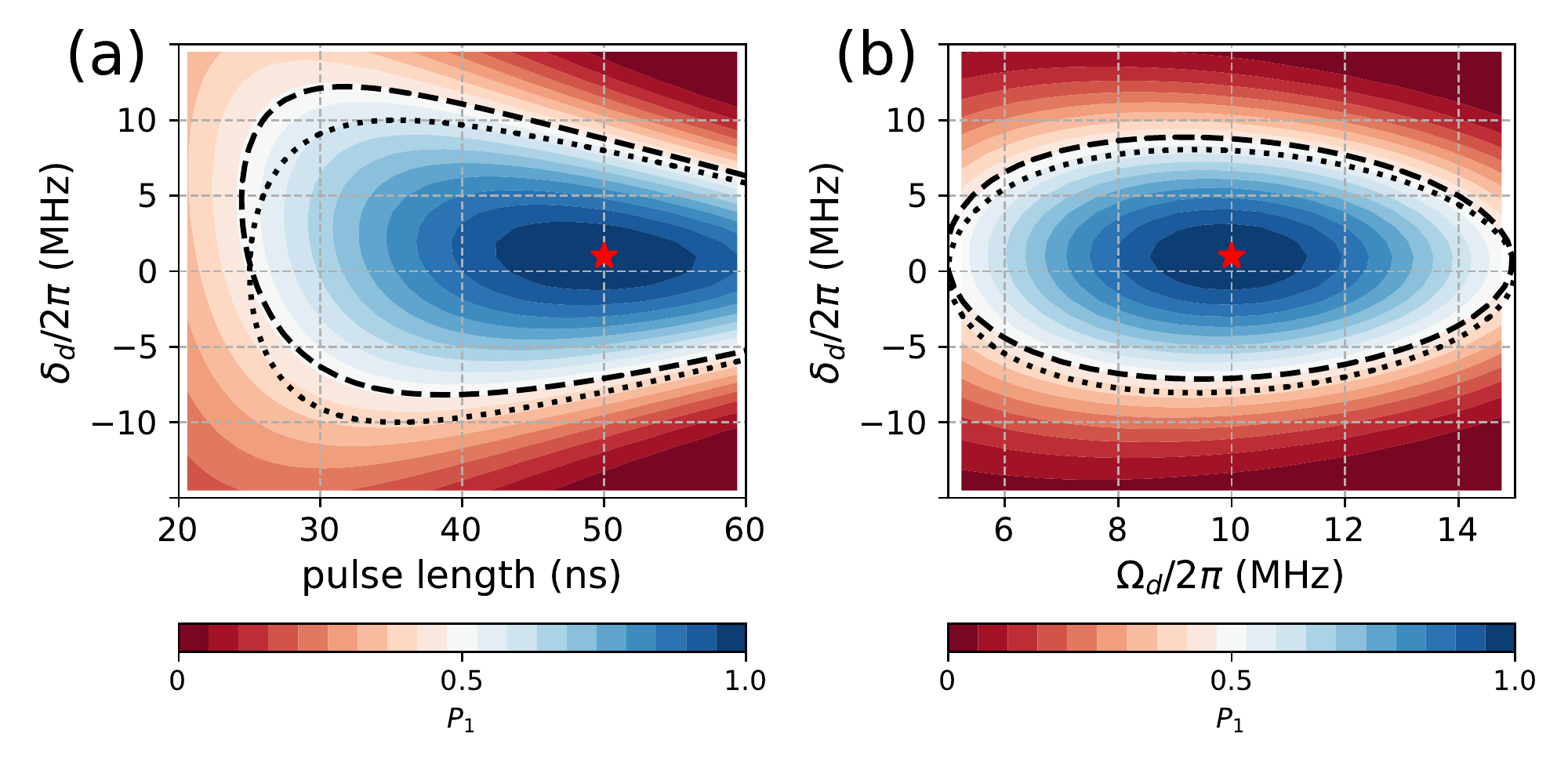}
\end{center}
\caption{Flexibility of $\sqrt{X}$ rotations. (a) Population in state $|1\rangle$ (i.e., $P_{1}$ at
the end of the pulse) versus the drive detuning and the pulse length of square pulses with the qubit prepared in
state $|0\rangle$. Here, the drive strength is $10\,\rm MHz$ and the drive detuning at the idle
point is $-100\,\rm MHz$. The red star indicates the optimal parameter set for implementing X
rotations, while the dashed and dotted lines indicate the available parameter sets for
implementing $\sqrt{X}$ rotations based on numerical simulations and analytical expression
in Eq.~(\ref{eq5}), respectively. (b) same as in (a), instead showing $P_{1}$ versus the drive detuning
and drive amplitude with the fixed gate length of $50\,\rm ns$.}
\label{fig2}
\end{figure}

Generally, as shown in Fig.~\ref{fig1}(c), by tuning the qubit on-resonance with
the global drive, single-qubit X rotations can be achieved. However, we note that since
the initial Bloch vector is slightly tilted, as shown in Fig.~\ref{fig1}(b), a small drive
detuning $\delta_{d}=|\Omega_{d}^{2}/\Delta_{d}|$ is needed for enabling ideal X rotations
with respect to the initial Bloch vector defined by Eq.~(\ref{eq3}). Thus, according to Z control
pulses, X rotations can be realized by tuning the qubit from the idle point to the working
point with a small overshoot \cite{Barends2019}. This fact is further illustrated by the
results shown in Fig.~\ref{fig2}. By initializing the qubit in state $|0\rangle$ and
using square pulses (results with cosine-decorated square pulses can be found in Appendix~\ref{A}),
Figure~\ref{fig2}(a) shows populations $P_{1}$, i.e., the population in
state $|1\rangle$ at the end of the applied pulse, versus the drive detuning $\delta_{d}$
and the pulse length. Here, the drive amplitude is $10\,\rm MHz$ and the detuning at the
idle point is $-100\,\rm MHz$. Similarly, given a fixed pulse length of $50\,\rm ns$, Figure~\ref{fig2}(b)
shows $P_{1}$ versus $\delta_{d}$ and $\Omega_{d}$. The optimal parameters for X rotations are
indicated by the red stars. Indeed, we find that a small frequency overshoot is needed
for X rotations.

In the present work, note that choosing $\sqrt{X}$ gates as the native gates
could simplify the tune-up procedure of single-qubit gate operations. This is because:

(i) arbitrary single-qubit rotations can be generated by two $\sqrt{X}$ gates
and three Z gates \cite{McKay2017}, i.e., $Z_{\phi1}-\sqrt{X}-Z_{\phi2}-\sqrt{X}-Z_{\phi3}$,
with $Z_{\phi}\equiv\exp[-i\phi Z/2]$;

(ii) compared with the native X gate, the implementation
of $\sqrt{X}$ gates does not pose stringent requirements on the on-resonance condition, i.e.,
even the qubit is slightly off-resonance with the drive, $\sqrt{X}$ gates can
still be achieved. This can be captured by the analytical
expression of Rabi oscillations for the two-level system initialized in state $|0\rangle$, i.e., Rabi's formula
\begin{equation}
\begin{aligned}\label{eq5}
P_{1}(t)=\frac{\Omega_{d}^{2}}{\Omega_{d}^{2}+\Delta_{d}^2}\sin^2\left[\frac{t}{2}\sqrt{\Omega_{d}^{2}+\Delta_{d}^2}\right].
\end{aligned}
\end{equation}
From Eq.~(\ref{eq5}), implementing $\sqrt{X}$ gates requires $P_{1}=1/2$ at the end of the applied
pulse, giving rise to the relations among the pulse length, the drive detuning $\Delta_{d}$, and the drive
amplitude $\Omega_{d}$, as illustrated by the dotted lines of Fig.~\ref{fig2}. Accordingly, the
results based on numerical simulations are also presented, as indicated by the dashed line
of Fig.~\ref{fig2}. Note here that the derivation of the analytical equation ignores
the slight tilt at the idle point, and this explains the discrepancy between the analytical
and numerical results. Both the analytical and numerical results show that compared to X rotations, the
available parameter ranges of $\sqrt{X}$ rotations can provide great flexibility in its
tune-up procedure.

Generally, due to the flexibility of $\sqrt{X}$ rotations, for tuning-up $\sqrt{X}$
gates, the above-mentioned overshoot can be ignored. In the next subsection, we will
show that following this way, given a fixed drive detuning, $\sqrt{X}$ gates can be realized
by only optimizing the ramp times of control pulses, as suggested by Fig.~\ref{fig2}(a).
Meanwhile, in large-scale quantum systems with multiplexed control, this flexibility can
be the most encouraging advantage as to mitigate single-qubit gate error due to
stray coupling between qubits and to compensate for the non-uniformity of qubit
parameters. This will be discussed in detail in Sec.~\ref{SecIV}.

\subsection{Baseband control of qubit with fast-adiabatic ramps}\label{SecIIB}

\begin{figure}[tbp]
\begin{center}
\includegraphics[width=4cm,height=4cm]{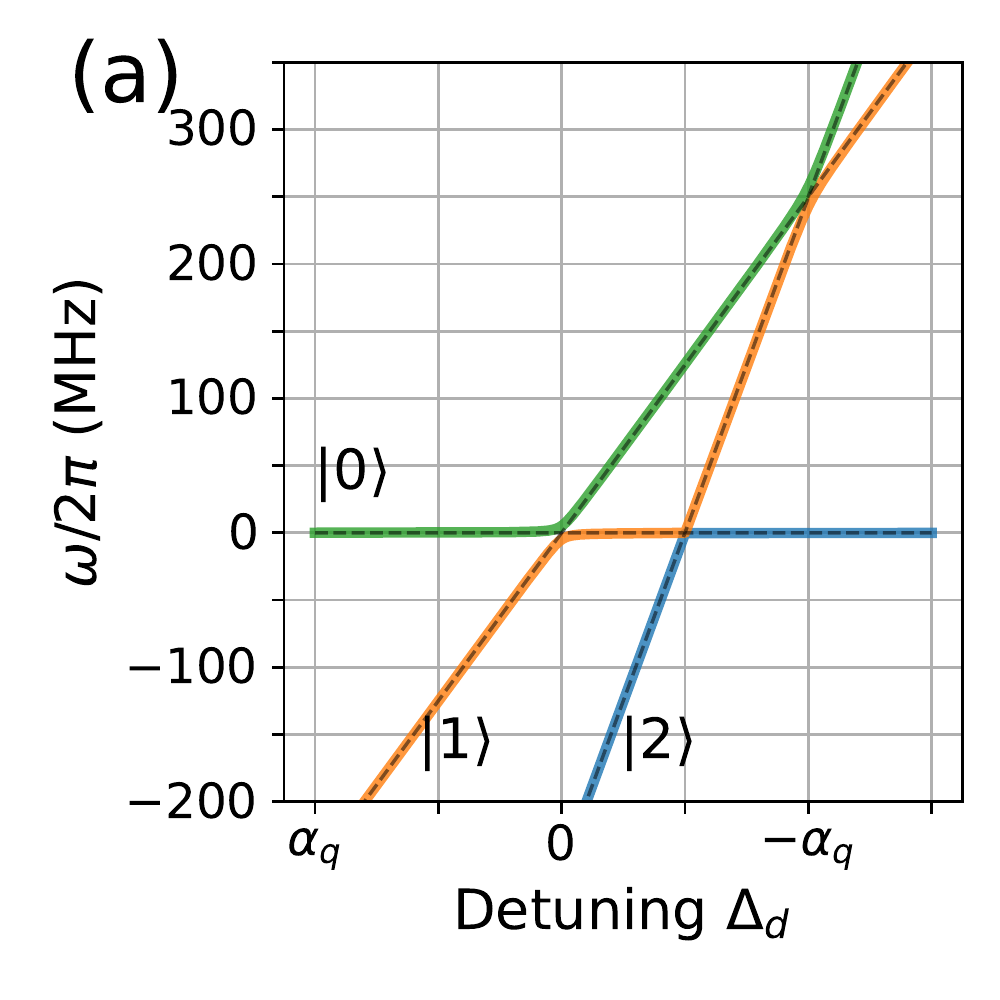}
\includegraphics[width=4cm,height=4cm]{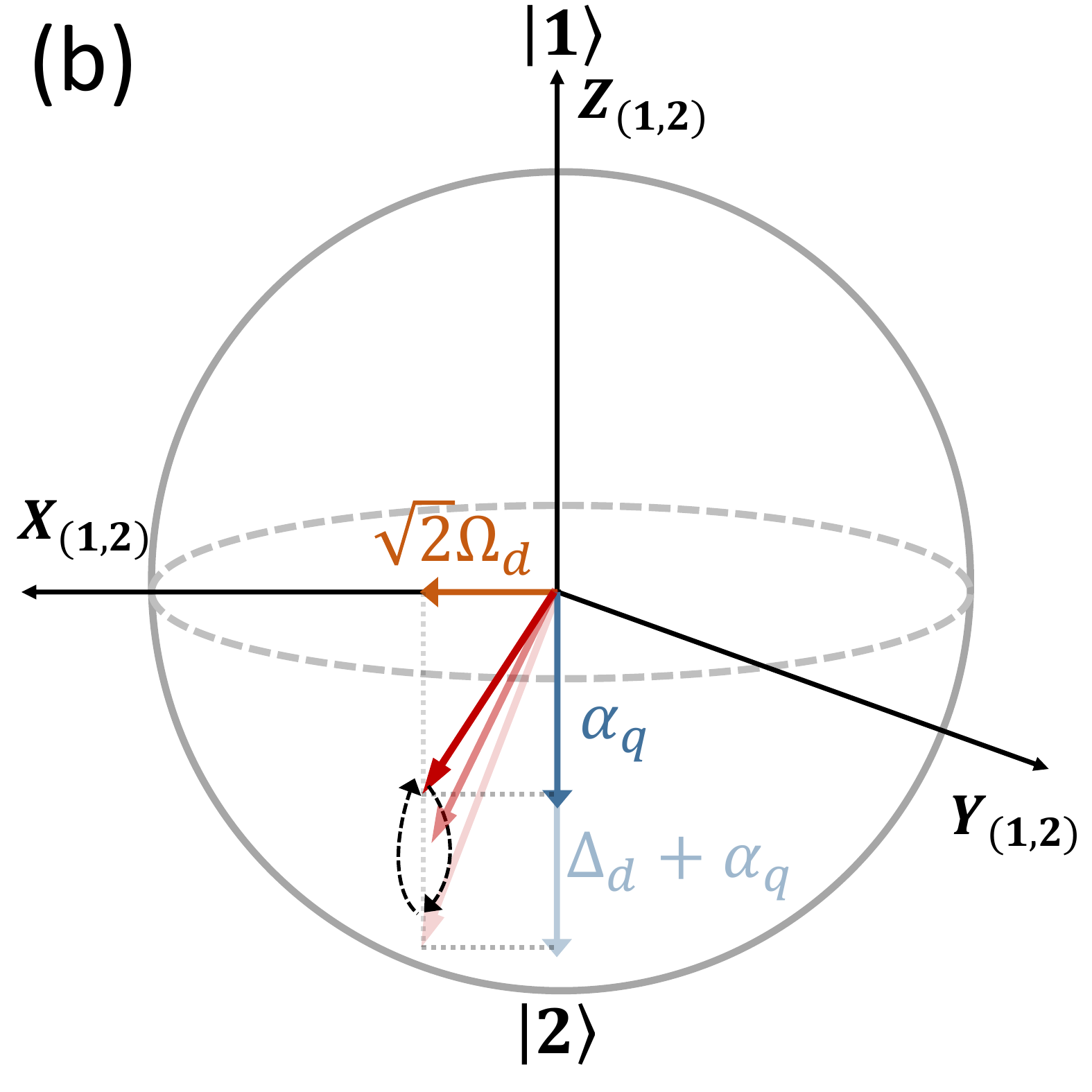}
\end{center}
\caption{Leakage out of the computational subspace. (a) Energy spectrum (solid lines) versus the drive detuning in the
rotating frame corresponding to the global drive. The dashed lines denote the bare energy levels (i.e., spectrum
without the drive). Here, the used system parameters are qubit
anharmonicity $\alpha_{q}/2\pi=-250\,\rm MHz$, drive frequency $\omega_{d}/2\pi=6.1\,\rm GHz$, and drive
amplitude $\Omega_{d}/2\pi=10\,\rm MHz$. (b) Bloch vector for the leakage space spanned by
states $\{|1\rangle,|2\rangle\}$. The strength of the coupling between states $|1\rangle$
and $|2\rangle$ is $\sqrt{2}\Omega_{d}$. At the idle point and in the leakage space, the
drive detuning is $\Delta_{d}+\alpha_{q}$, while at the working point, the
detuning is $\alpha_{q}$. During single-qubit X rotations, the Bloch vector in this leakage
space varies according to the drive detuning. In the present work, to avoid possible leakage
during qubit control, the qubit idle frequency is far detuned below the drive frequency.}
\label{fig3}
\end{figure}

In the above discussion, the single-qubit baseband control is discussed for an
ideal two-level system. Nevertheless, for practical superconducting qubits, such
as the transmon qubit, the weak qubit anharmonicity makes single-qubit gate operations
particularly prone to leakage outside the qubit subspace. For one such baseband control transmon
qubit, which is driven by an always-on global drive, the system Hamiltonian is (hereafter, transmon qubits
are modeled as anharmonic oscillators \cite{Koch2007})

\begin{equation}
\begin{aligned}\label{eq6}
H_{q}=\omega_{q}a_{q}^{\dagger}a_{q}+\frac{\alpha_{q}}{2}a_{q}^{\dagger}a_{q}^{\dagger}a_{q}a_{q}
+\frac{\Omega_{d}}{2}(a_{q}^{\dagger}e^{-i\omega_{d}t}+a_{q}e^{+i\omega_{d}t}).
\end{aligned}
\end{equation}
Here, $a_{q}\,(a_{q}^{\dagger})$ is the annihilation (creation) operator.
Figure~\ref{fig3}(a) shows the energy spectrum of the driven qubit versus the
drive detuning with qubit anharmonicity $\alpha_{q}/2\pi=-250\,\rm MHz$,
drive frequency $\omega_{d}/2\pi=6.1\,\rm GHz$, and drive
amplitude $\Omega_{d}/2\pi=10\,\rm MHz$. One can find that due to the
global drive, there exits an off-resonance coupling
between states $|2\rangle$ and $|1\rangle$, which can cause leakage to state $|2\rangle$
when performing X rotations, i.e., biasing the qubit from its idle point to the
working point. Note that there exist two leakage channels, caused by the $|1\rangle\leftrightarrow|2\rangle$
and $|0\rangle\leftrightarrow|2\rangle$ interactions, as shown in Fig.~\ref{fig3}(a). Since the $|0\rangle\leftrightarrow|2\rangle$
interaction involves second-order processes, generally, the induced leakage error is far smaller than
that of the $|1\rangle\leftrightarrow|2\rangle$ interaction. Here, we thus mainly focus
on the leakage error caused by the $|1\rangle\leftrightarrow|2\rangle$ interaction.
However, we note that in our numerical analysis, all the leakage channels, including
the $|0\rangle\leftrightarrow|2\rangle$ interaction, are taken into consideration.
While this leakage issue can be addressed by using the DRAG scheme
in the traditional microwave control setup, this scheme cannot be directly utilized
for the baseband flux control setup, as here only Z control
is available.

Additionally, in principle, at the idle point, qubits could be far detuned
above or below the frequency of the always-on drive. However,
to avoid possible leakage error during qubit control, such as gate operations, qubit initialization,
and readout, we prefer to bias the qubit away from the
harmful avoid crossing caused by coupling between states $|1\rangle$ and $|2\rangle$, as
shown in Fig.~\ref{fig3}(a). In this way, during the baseband-controlled gate operations,
the qubit system will not sweep through or operate nearby this harmful avoid crossing, generally
allowing the suppression of the leakage to state $|2\rangle$. Considering
this fact, hereafter, we consider biasing the qubit below the drive frequency at
the idle point. Even in the setting, during Z-controlled single-qubit gate
operations, leakage error can still occur due to the non-adiabatic error, as shown
in Fig.~\ref{fig3}(b). In the following, we will consider using a
fast-adiabatic control scheme for suppressing the leakage further.

\begin{figure}[tbp]
\begin{center}
\includegraphics[keepaspectratio=true,width=\columnwidth]{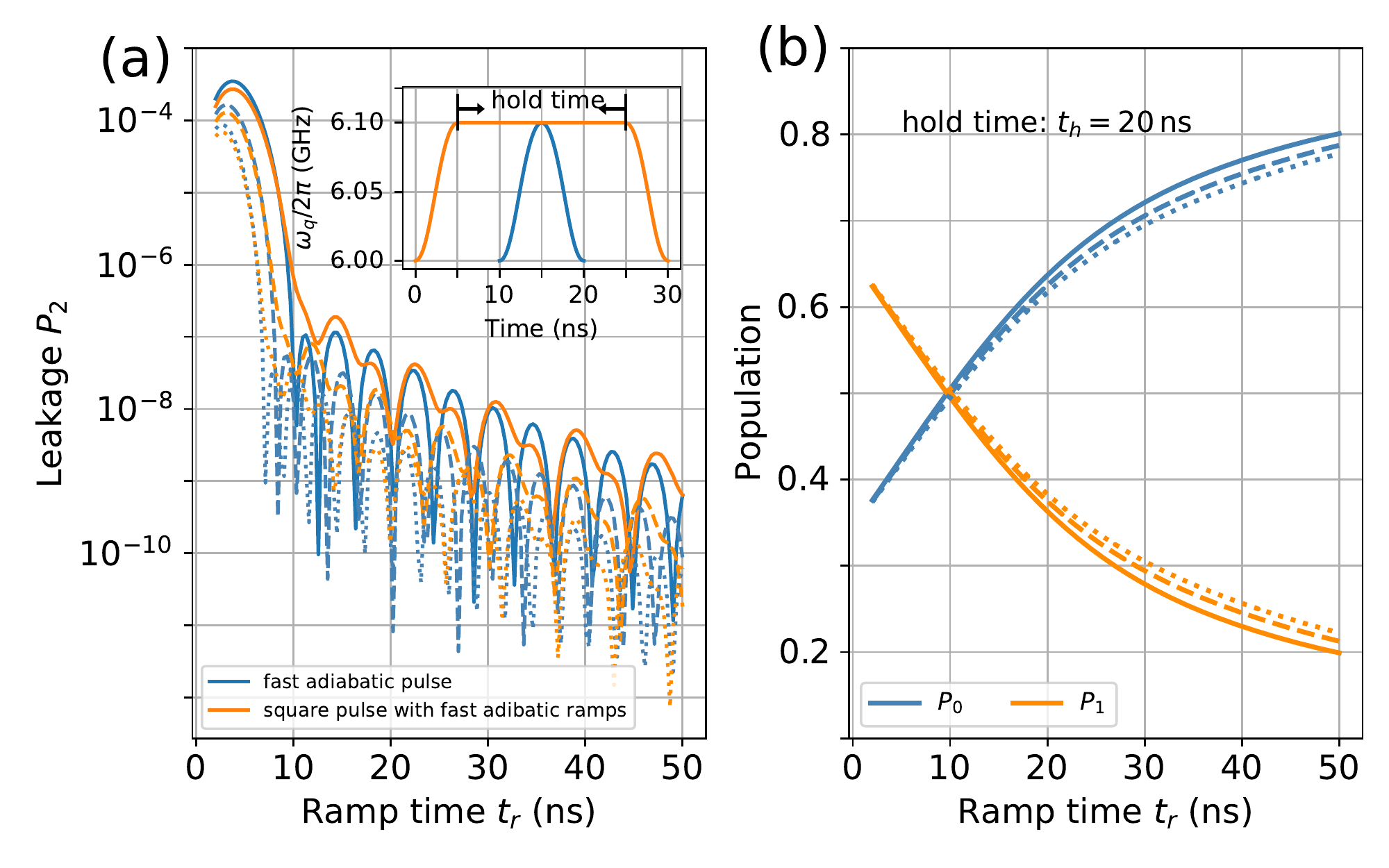}
\end{center}
\caption{Minimizing leakage out of the computational subspace for performing Z-controlled X rotations.
(a) Leakage as a function of the ramp time for the transmon qubit initialized in
state $|1\rangle$ with the anharmonicity $\alpha_{q}/2\pi=\{-200,\,-250,\,-300\}\rm MHz$ (
denoted by the solid line, dashed line, and dotted line, respectively). Inset shows
the typical fast-adiabatic pulse and the fast-adiabatic flat-top pulse (i.e., square pulse with
fast-adiabatic ramps) for controlling the qubit frequency from the idle point ($6.0\,\rm GHz$) to
the working point ($6.1\,\rm GHz$). The other system parameters
are: the hold time (i.e., the pulse length of the flat part) $t_{h}=20\,\rm ns$,
drive frequency $\omega_{d}/2\pi=6.1\,\rm GHz$, and drive amplitude $\Omega_{d}/2\pi=10\,\rm MHz$.
(b) Same as in (a), instead showing the population in states $|0\rangle$ and $|1\rangle$
versus the ramp time of the fast-adiabatic flat-top pulse.}
\label{fig4}
\end{figure}

As shown in Fig.~\ref{fig3}(b), the leakage error occurs when one non-adiabatically varies
the driving detuning. Considering that coherence times of superconducting qubits are still limited, our target
is to find a good flux control pulse, thus the non-adiabatic error is suppressed
while maintaining a fast operation speed. Fortunately, this issue has already been addressed
successfully by using a fast-adiabatic scheme introduced in Ref.~\cite{Martinis2014b}. Within
the scheme, optimal control pulses can be obtained for minimizing non-adiabatic errors for any pulse
longer than the chosen pulse length. However, we note that the original scheme
only addresses the non-adiabatic error in the pulse ramps. Thus, here, to address the leakage
issue in our setting, we consider using a square control pulse with optimal fast-adiabatic ramps, which is obtained
following the fast-adiabatic scheme \cite{Martinis2014b} (see Appendix~\ref{B} for
details). The inset of Fig.~\ref{fig4}(a) shows the optimal ramp pulse (solid blue line), which is used for
generating our target control pulse with a flat middle part and fast-adiabatic
ramps (solid orange line). Hereafter, we refer to this pulse as the fast-adiabatic flat-top pulse.

Here, we turn to evaluate the efficiency of the proposed fast-adiabatic flat-top pulse. We consider
that the qubit idle frequency is $6.0\,\rm GHz$ and during the implementation of
single-qubit X rotations, the drive detuning $\Delta_{d}$ varies from the idle
point at $-100\,\rm MHz$ to the work point at $0\,\rm MHz$ and then coming
back, according to the fast-adiabatic flat-top pulse. By initializing the qubit in state $|1\rangle$,
Figure~\ref{fig4}(a) shows the population leakage to state $|2\rangle$ as a function of ramp times
with the hold time of 20 ns and the qubit anharmonicities $\alpha_{q}/2\pi=\{-200,\,-250,\,-300\}\rm MHz$.
For easy comparison, the results for applying only the fast-adiabatic pulse are also presented. One
can find that by using the fast-adiabatic flat-top pulse, the leakage can be suppressed
below $10^{-6}$ for ramp times longer than 10 ns, and inserting a square pulse in the
fast-adiabatic pulse does not change the efficiency of the original fast-adiabatic
scheme. In Fig.~\ref{fig4}(b), we also show the populations in $|0\rangle$
and $|1\rangle$ versus the ramp times. Additionally, Appendix~\ref{B} presents further
results for different drive strengths.

From the results shown in Fig.~\ref{fig4}, one can find that $\sqrt{X}$ rotations can be
realized with the ramp time at about 10 ns, giving rise to the total pulse length of about 30 ns.
Meanwhile, same as the case for two-level systems (in Sec.\ref{SecIIA}), here, single-qubit Z rotations
can be easily implemented by controlling the time delay between Z pulses. Therefore, we could
reasonably expect that with the help of the fast-adiabatic scheme, fast-speed single-qubit
operations could be achieved with low leakage errors (we will evaluate the single-qubit gate
performance in detail in the following section).

\subsection{Dephasing due to fluctuations in the drive amplitude}\label{SecIIC}

\begin{figure}[tbp]
\begin{center}
\includegraphics[keepaspectratio=true,width=\columnwidth]{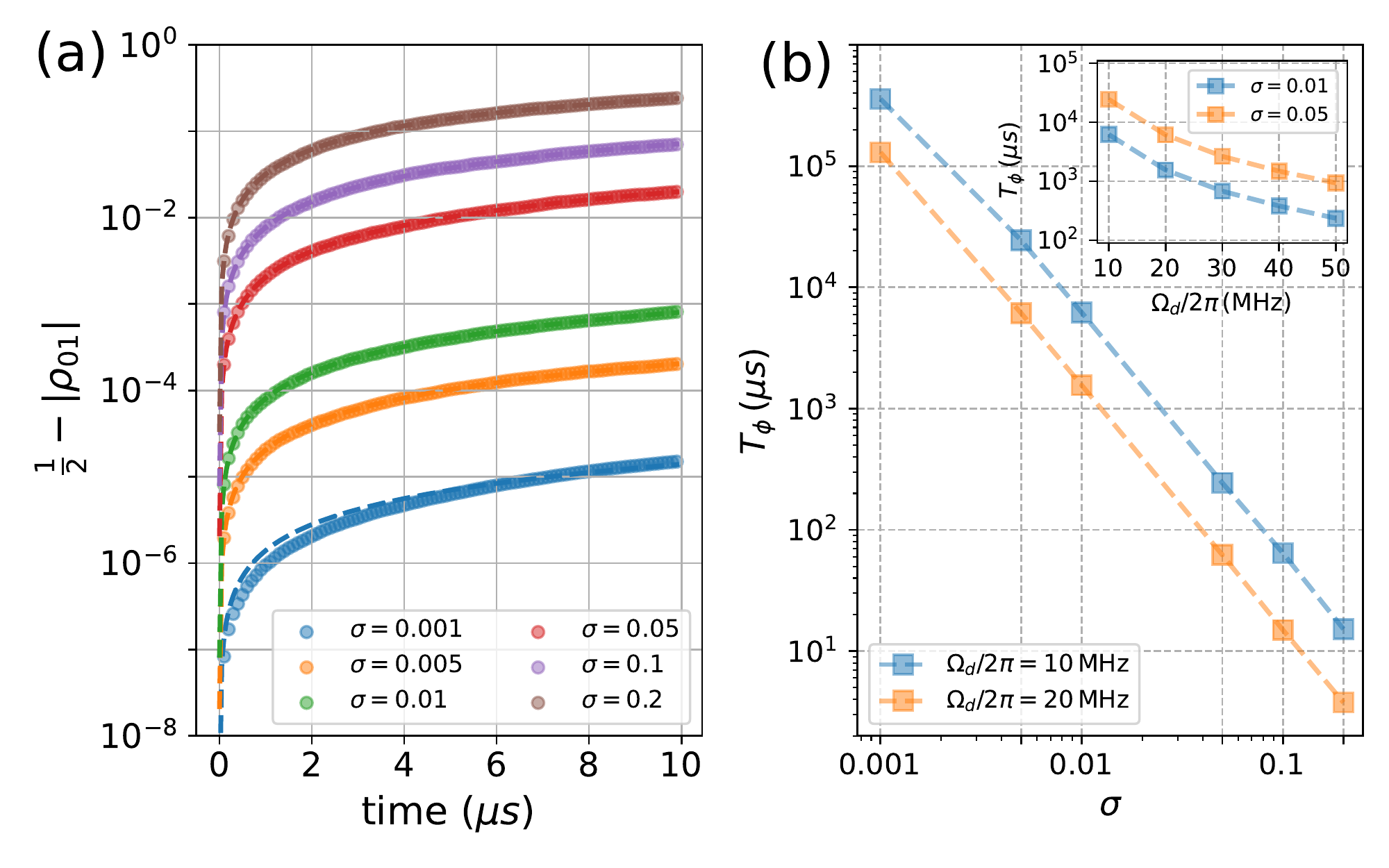}
\end{center}
\caption{Qubit dephasing due to the fluctuations in the amplitude of the always-on
drive. (a) Time evolution of the magnitudes of the averaged off-diagonal matrix
element, i.e., $|\langle\rho_{01}(t)\rangle|$, for the qubit initialized in
state $(|0\rangle+|1\rangle)/\sqrt{2}$. Here, we assume that the drive
amplitudes ($\Omega_{d}/2\pi=10\,\rm MHz$) subject to amplitude-dependent
Gaussian noise, i.e., $\textsl{N}(0,\sigma)$, and 2000 realizations
of noise are used for obtaining $|\langle\rho_{01}(t)\rangle|$. The dashed lines are
exponential fits $[1-\exp(-t/T_{\phi})]/2$, giving rise to the dephasing
time $T_{\phi}$. (b) Dephasing time versus the noise variance. For given
noise variances, the inset shows the dephasing time versus the drive amplitude. Here, the
other parameters used are: $\Delta_{d}/2\pi=-100\,\rm MHz$
and $\alpha_{q}/2\pi=-250\,\rm MHz$.}
\label{fig5}
\end{figure}

Within the introduced baseband control setup, at the idle point, the global
always-on drive acts as an off-resonance drive and can induce ac-Stark
frequency shifts on the qubits. For the two-level system studied
in Sec.\ref{SecIIA}, the shift is given
as $\delta\omega=\Delta-\Delta_{d}\approx \Omega_{d}^{2}/(2\Delta_{d})$,
while, taking the higher energy levels of the transmon qubit into
consideration, the shift is \cite{Schneider2018}
\begin{equation}
\begin{aligned}\label{eq7}
\delta\omega\approx \frac{\alpha_{q}\Omega_{d}^{2}}{2\Delta_{d}(\Delta_{d}+\alpha_{q})}.
\end{aligned}
\end{equation}
From Eq.~(\ref{eq7}), the shift has a quadratic-dependent on the drive amplitude, making
the qubit frequency more susceptible to possible amplitude noise. Therefore, fluctuations in the drive
amplitude can cause qubit dephasing, which has been recently observed in superconducting
qubits \cite{Wei2022}.

Here, to numerically study the amplitude-fluctuation-induced qubit dephasing, we consider
an amplitude-dependent noise, i.e., amplitude fluctuations are proportional to
the amplitudes. By assuming the drive subject to zero-mean Gaussian
noise, i.e., $\textsl{N}(0,\sigma)$, we numerically simulate the time evolution of the
off-diagonal matrix element $\rho_{01}(t)$ for the qubit initialized in
state $(|0\rangle+|1\rangle)/\sqrt{2}$. After averaging $\rho_{01}(t)$ over 2000
trajectories (i.e., realizations of noise), the magnitudes of the off-diagonal matrix element display a clear
exponential decay, as shown in Fig.~\ref{fig5}(a). Here, the evolution time
is $10\,\mu s$, the other used parameters are: $\Delta_{d}/2\pi=-100\,\rm MHz$,
$\Omega_{d}/2\pi=10\,\rm MHz$, and $\alpha_{q}/2\pi=-250\,\rm MHz$, giving
rise to $\delta\omega/2\pi\approx-0.36\,\rm MHz$. By fitting the decay curves
to $[1-\exp(-t/T_{\phi})]/2$, Figure~\ref{fig5}(b) shows the dephasing time $T_{\phi}$
versus the noise variance $\sigma$. Here, we also show the results for
$\Omega_{d}/2\pi=20\,\rm MHz$. Additionally, in the inset, we further
show the dephasing times versus the drive amplitudes.

From the results shown in Fig.~\ref{fig5}(b), and given the typical noise variance of $1\%$, we
can conclude that the amplitude-noise induced dephasing can be safely neglected by detuning the
qubit far from the drive frequency. Meanwhile, we note that to ensure high-fidelity gate
operations within sub-100 ns, the drive amplitude itself should be larger than $10\,\rm MHz$.
Finally, we note that besides the qubit dephasing, when the always-on drive is shared
by multiple qubits, there are two additional potential issues related to the drive and its fluctuation:

(i) qubit decoherence due to the
excess quasiparticles (QPs) \cite{Catelani2011}. Previous works have demonstrated that QPs can be
injected into a qubit by applying a high-power microwave pulse resonance
with its readout resonator \cite{Vool2014,Wang2014}. However, at the present setting, the strength of the
always-on drive is far smaller than that of the former case, we thus
expect that the contribution of the always-on drive to the
excess QPs is negligible.

(ii) when the always-on shared (global) drive is shared by
multiple qubits, fluctuations in the drive can result in errors on
multiple qubits. At first glance, this can cause correlated errors.
However, as discussed in Ref.~\cite{Fowler2014}, provided the fluctuation is small and
quantum error correction is frequent, the fluctuation in the shared drive can result in a
correlated probability of error on multiple qubits, but the errors themselves will not be
correlated. Additionally, as suggested in Fig.5(b) and Eq.(7), by increasing the qubit-drive
detuning, the fluctuation-induced dephasing (error) can be heavily suppressed at the system idle time. And
during single-qubit gate operations, with the currently available control electronics, single-qubit
gates with gate errors below 0.0001 have been demonstrated \cite{Somoroff2021,Li2023}. Thus, we argue that
fluctuations in the shared drive itself can indeed cause errors, but probably a
rather small one ($<0.0001$). In this case, the control-noise-induced error may still be
handled by the error correction schemes \cite{Fowler2014}.

\subsection{Impact of the always-on drive on the qubit readout}\label{SecIID}

\begin{figure}[tbp]
\begin{center}
\includegraphics[keepaspectratio=true,width=\columnwidth]{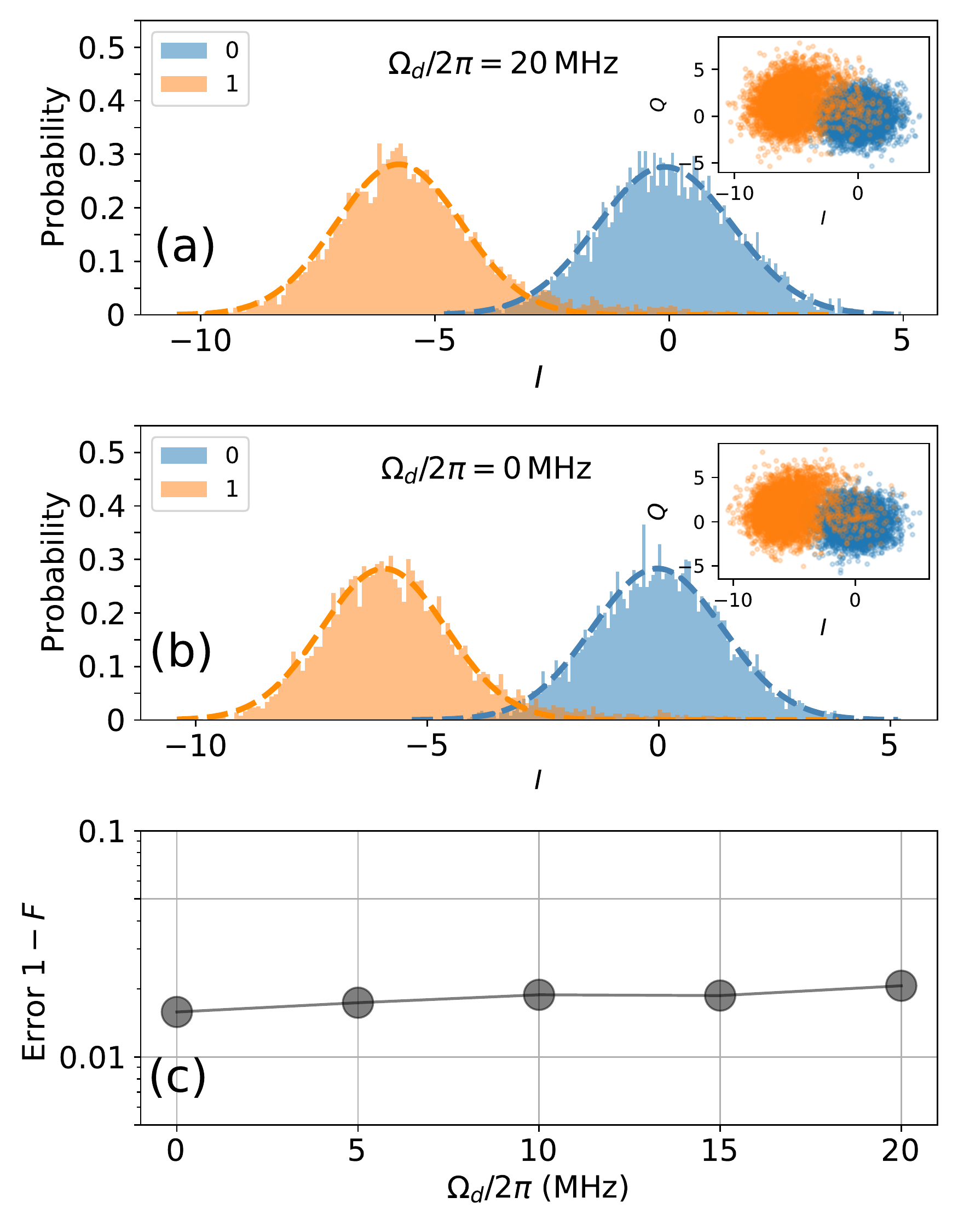}
\end{center}
\caption{Qubit dispersive readout with the presence of the always-on drive.
(a) Histograms of the integrated readout quadrature for the qubit prepared in
states $|0\rangle$ (blue) and $|1\rangle$ (orange). The dashed blue line and
dashed orange line denote the Gaussian fits of the histograms for states $|0\rangle$
and $|1\rangle$, respectively. The intersection point of the two fitted distributions
gives the state-decision threshold. The inset shows the IQ scatter plot of the
integrated readout quadrature. (b) Same as in (a), instead showing the results
without the always-on drive. (c) Readout error $1-F$ as a function of the drive
strength.}
\label{fig6}
\end{figure}

As mentioned in Sec.~\ref{SecI} and Sec.~\ref{SecIIA}, due to the presence
of the always-on drive, in this work, the microwave dressed states are defined as the
computational states. Here, since the available control over the always-on
drive is limited, the previous method \cite{Zhao2022,Huang2021}, in which the dressed state is first mapped back to
the corresponding bare state, and then the traditional dispersive readout is employed for
inferring the qubit information \cite{Wallraff2005}, cannot be directly utilized. However, as
discussed in Sec.~\ref{SecIIA}, when the drive detuning is far larger than the
drive amplitude, i.e., $|\Delta_{d}|\gg\Omega_{d}$, the difference between dressed states and
bare states can be neglected. Therefore, we expect that by keeping a large ratio of the
drive detuning to the drive amplitude, the qubit information can be directly inferred
using the traditional dispersive readout scheme.

To explore the possible impact of the always-on drive on the qubit dispersive readout, we numerically
simulate the system dynamics during the dispersive readout. By applying a 250-ns square readout pulse with
frequency $\omega$ and amplitude $\Omega$ to the readout resonator with decay rate $\kappa$, the
full system dynamics are governed by the Hamiltonian
\begin{equation}
\begin{aligned}\label{eq8}
H_{\rm read}=&H_{q}+\omega_{r}a_{r}^{\dagger}a_{r}+g(a_{q}^{\dagger}a_{r}+a_{q}a_{r}^{\dagger})
\\&+\frac{\Omega}{2}(a_{r}^{\dagger}e^{-i\omega t}+a_{r}e^{+i\omega t}),
\end{aligned}
\end{equation}
where $H_{q}$ denotes the qubit Hamiltonian given in Eq.~(\ref{eq6}), $\omega_{r}$ is the frequency of
the readout resonator, $a_{r}\,(a_{r}^{\dagger})$ is the annihilation (creation) operator of
the resonator, and $g$ denotes the strength of the qubit-resonator coupling. In this following,
the qubit information is encoded into single quadrature, i.e., $I$-quadrature, by choosing the readout
frequency to be $\omega=(\omega_{r0}+\omega_{r1})/2$ \cite{Wallraff2005}. Here, $\omega_{r0}$ and $\omega_{r1}$ denote the
dressed resonator frequencies with the qubit in states $|0\rangle$ and $|1\rangle$, respectively. The other
system parameters are: $\omega_{q}/2\pi=6.0\,\rm GHz$,
$\alpha_{q}/2\pi=-250\,\rm MHz$, $\omega_{d}/2\pi=6.1\,\rm GHz$, $\omega_{r}/2\pi=5.0\,\rm GHz$,
$g/2\pi=100\,\rm MHz$, $\kappa/2\pi=5\,\rm MHz$, and $\Omega/2\pi=7\,\rm MHz$.

According to Eq.~(\ref{eq8}), we simulate the system dynamics based on solving the stochastic master
equation \cite{Johansson2012}. Then,
following Ref.~\cite{Walter2017}, we further caulate the integrated readout quadrature with an optimal
weight function (see Appendix~\ref{C} for details). With 5000 repetitions of the simulation for each qubit basis
state, i.e., $|0\rangle$ and $|1\rangle$, Figure~\ref{fig6}(a) shows the two histograms of the
integrated readout quadrature with the qubit prepared in states $|0\rangle$ and $|1\rangle$,
respectively. Here, the drive magnitude is $20\,\rm MHz$. For easy comparison, we also present
the result for the global drive is absent, as shown in Fig.~\ref{fig6}(b).

Fitting the histograms to Gaussian functions gives the state-decision threshold at the
intersection point of the two fitted distributions. Accordingly, the readout fidelity
can be calculated as $F=1-[P(0|1)+P(1|0)]/2$, where $P(0|1)$ ($P(1|0)$) denotes the error probability
that the qubit initialized in state $|1\rangle$ ($|0\rangle$) is identified as in
state $|0\rangle$ ($|1\rangle$). Accordingly, Figure~\ref{fig6}(c) shows the readout error $1-F$ versus the drive strength. One can
find that when increasing the drive amplitude from $0$ to $20\,\rm MHz$, while the error
shows an upward trend, the increased error is below $1\%$. Moreover, the upward trend also suggests
that by further increasing the ratio $|\Delta_{d}|/\Omega_{d}$, the increased error should be
heavily suppressed.

\section{An Application in qubit architectures with tunable coupling}\label{SecIII}

\begin{figure}[tbp]
\begin{center}
\includegraphics[keepaspectratio=true,width=\columnwidth]{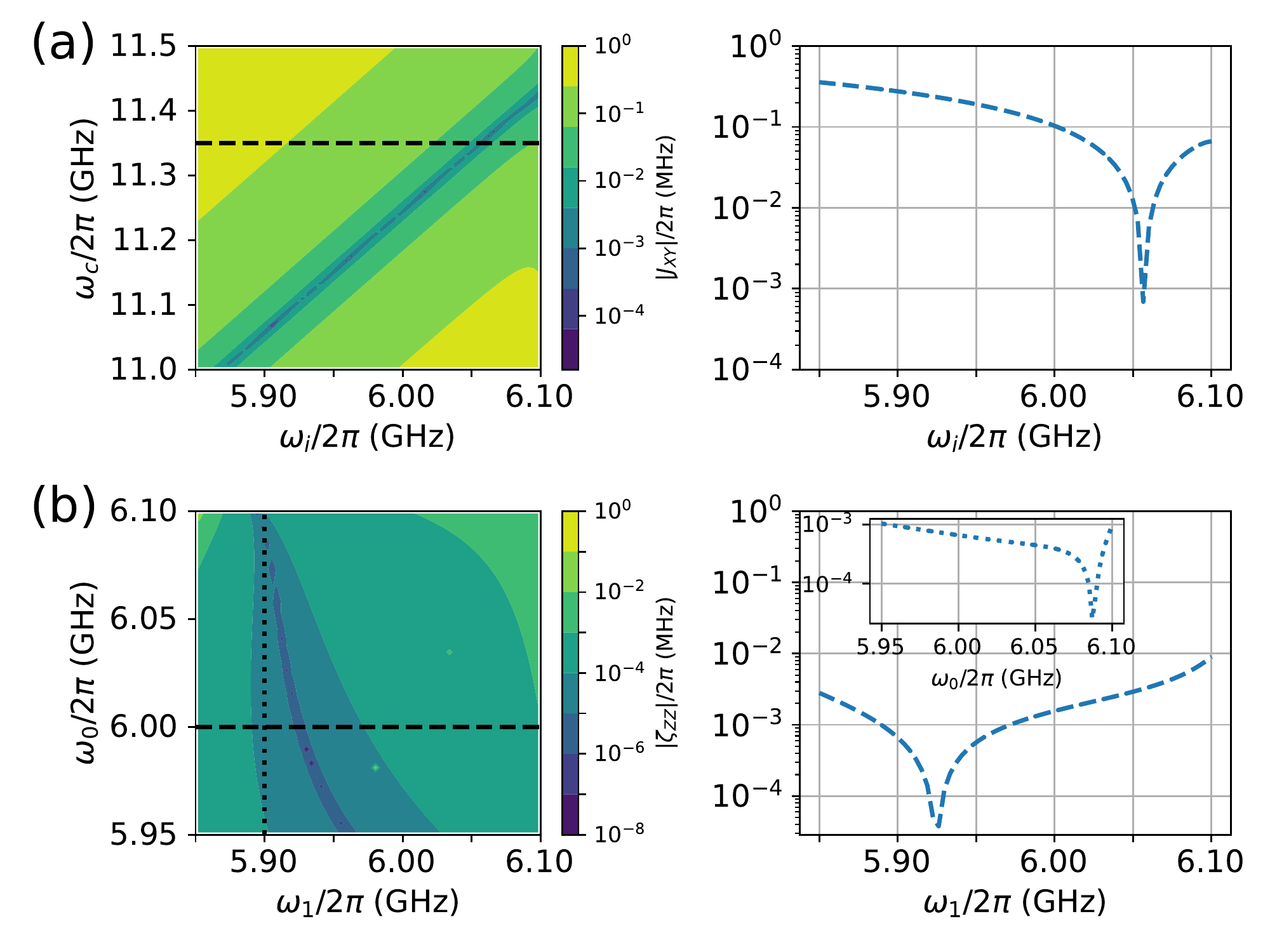}
\end{center}
\caption{Residual coupling with varying qubit frequency. (a) Left: residual resonance XY coupling
versus the qubit frequency and the coupler frequency. Horizontal cut through (Left) denotes the
result plotted in (Right), i.e., XY coupling versus the qubit frequency with the coupler
frequency fixed at $\omega_{c}/2\pi=11.35\,\rm GHz$. (b) Left: residual ZZ coupling versus the
frequencies ($\omega_{0}$ and $\omega_{1}$) of the two qubits with the coupler frequency
fixed at $11.35\,\rm GHz$. Horizontal cut through (Left) denotes the
result plotted in (Right), i.e., ZZ coupling versus frequency of $Q_{1}$ with the $Q_{0}$'s
frequency fixed at $\omega_{0}/2\pi=6.0\,\rm GHz$. Vertical cut through (Left) denotes the result plotted
in the inset of (Right), i.e., ZZ coupling versus the frequency of $Q_{0}$ with the $Q_{1}$'s
frequency fixed at $\omega_{1}/2\pi=5.9\,\rm GHz$.}
\label{fig7}
\end{figure}

Given the overview of the baseband control scheme, in this section, we will present the
application of this scheme in a qubit architecture with tunable coupling.  As depicted
in Fig.~\ref{fig1}(a), we consider that two frequency-tunable transmon qubit $Q_{0}$ and $Q_{1}$ are
coupled via a tunable coupler $Q_{c}$ (i.e., an auxiliary transmon qubit) and both
qubits are driven by an always-on global drive. After applying RWA, the system
Hamiltonian is given by

\begin{equation}
\begin{aligned}\label{eq9}
H=&\sum_{j=0,1,c}\big(\omega_{j}a_{j}^{\dagger}a_{j}+\frac{\alpha_{j}}{2}a_{j}^{\dagger}a_{j}^{\dagger}a_{j}a_{j}\big)
\\&+\sum_{\substack{k=0,1,c\\j\neq k}}g_{jk}(a_{j}a_{k}^{\dagger}+a_{j}^{\dagger}a_{k})
\\&+\sum_{i=0,1}\frac{\Omega_{d}}{2}(a_{i}^{\dagger}e^{-i\omega_{d}t}+a_{i}e^{+i\omega_{d}t}),
\end{aligned}
\end{equation}
where $\omega_{j}$ and $\alpha_{j}$ are the bare qubit frequency and the qubit anharmonicity
of $Q_{j}$, $q_{j}\,(q_{j}^{\dagger})$ is the associated annihilation (creation)
operator, and $g_{jk}$ denotes strength of the coupling between $Q_{j}$ and $Q_{k}$. Hereafter, the
system state is denoted by the notation $|Q_{0}Q_{c}Q_{1}\rangle$ and
the used system parameters are: the qubit anharmonicity $\alpha_{0}/2\pi=\alpha_{1}/2\pi=-250\,\rm MHz$,
the coupler anharmonicity $\alpha_{c}/2\pi=-200\,\rm MHz$, the direct qubit-qubit coupling
strength $g_{01}/2\pi=13\,\rm MHz$ (at $\omega_{0}/2\pi=\omega_{1}/2\pi=5.5\,\rm GHz$),
the qubit-coupler coupling strength $g_{0c}/2\pi=g_{1c}/2\pi=160\,\rm MHz$
(at $\omega_{0(1)}/2\pi=\omega_{c}/2\pi=5.5\,\rm GHz$), the drive
amplitude $\Omega_{d}/2\pi=10\,\rm MHz$, and the drive frequency $\omega_{d}/2\pi=6.1\,\rm GHz$.

Note here that the RWA is used for simplifying numerical simulation (otherwise, given the always-on
drive, Floquet methods could be employed here \cite{Huang2021,Shirley1965,Sambe1973,Petrescu2021}). However, in
the present two-qubit system with tunable coupling, the non-RWA terms in the original Hamiltonian (see Appendix~\ref{D}
for details) can significantly affect the effective coupling between qubits and can shift
the bare qubit frequency. Thus, here, considering non-RWA terms while
still working within the RWA formalism, we keep second-order corrections from the non-RWA
terms and find that with this correction (details on its derivation can be
found in Appendix~\ref{D}), the results agree well with the results without
applying the RWA. Accordingly, the corrections are taken into consideration
throughout the following discussion.

Before going into details of the baseband controlled gate operations, we give a few brief discussions of
the tunable coupling architecture. For performing gate operations in multiqubit systems, the key
benefit of the introduced tunable coupler is that the inter-qubit coupling strength can be tuned
off by biasing the coupler at a certain frequency point, i.e., zero-coupling point. However, the
zero-coupling point can change when the qubit is just biased slightly away from its idle point.
Fortunately, in the tunable coupling architecture, biasing
the qubit slightly away, generally, only causes a small increase in the residual inter-qubit
coupling. This can be found in Fig.~\ref{fig7}. Figure~\ref{fig7}(a) shows the strengths of the
residual resonance XY coupling versus the qubit frequency and the coupler frequency, while
Figure~\ref{fig7}(b) shows the residual ZZ coupling versus the frequencies of the two qubits with the
coupler frequency fixed at $11.35\,\rm GHz$. Here, the XY coupling and the ZZ coupling are
numerically calculated by the diagonalization of the Hamiltonian Eq.~(\ref{eq9}) in the rotating
frame defined by the always-on drive. To be more specific, the XY coupling is extracted as half the
energy difference between dressed eigenstates $|10\rangle$ and $|01\rangle$, while the ZZ coupling
is $\zeta_{zz}=(E_{11}-E_{10})-(E_{01}-E_{00})$. Here, $E_{ij}$ denotes the energy of
dressed eigenstate $|ij\rangle$, which is adiabatically connected to the
bare state $|i0j\rangle$ \cite{Ghosh2013}.

According to the above results, in the following discussion, we consider that at the
system idle point, the frequency of qubit $Q_{0}$ and qubit $Q_{1}$ are $\omega_{0}/2\pi=6.0\,\rm GHz$
and $\omega_{1}/2\pi=5.9\,\rm GHz$, respectively, and the frequency of coupler $Q_{c}$ is
$\omega_{c}/2\pi=11.35\,\rm GHz$. Therefore, the residual ZZ coupling is below $10\,\rm kHz$ at
the system idle point. Moreover, during the gate operations based
on slightly tuning qubit frequency, such as implementing single-qubit gates by tuning the qubit from its idle
point to the working point (e.g., at $6.1\,\rm GHz$), the residual inter-qubit ZZ coupling
can always be below $10\,\rm kHz$.

\subsection{Single-qubit gate operation}\label{SecIIIA}

\begin{figure}[tbp]
\begin{center}
\includegraphics[keepaspectratio=true,width=\columnwidth]{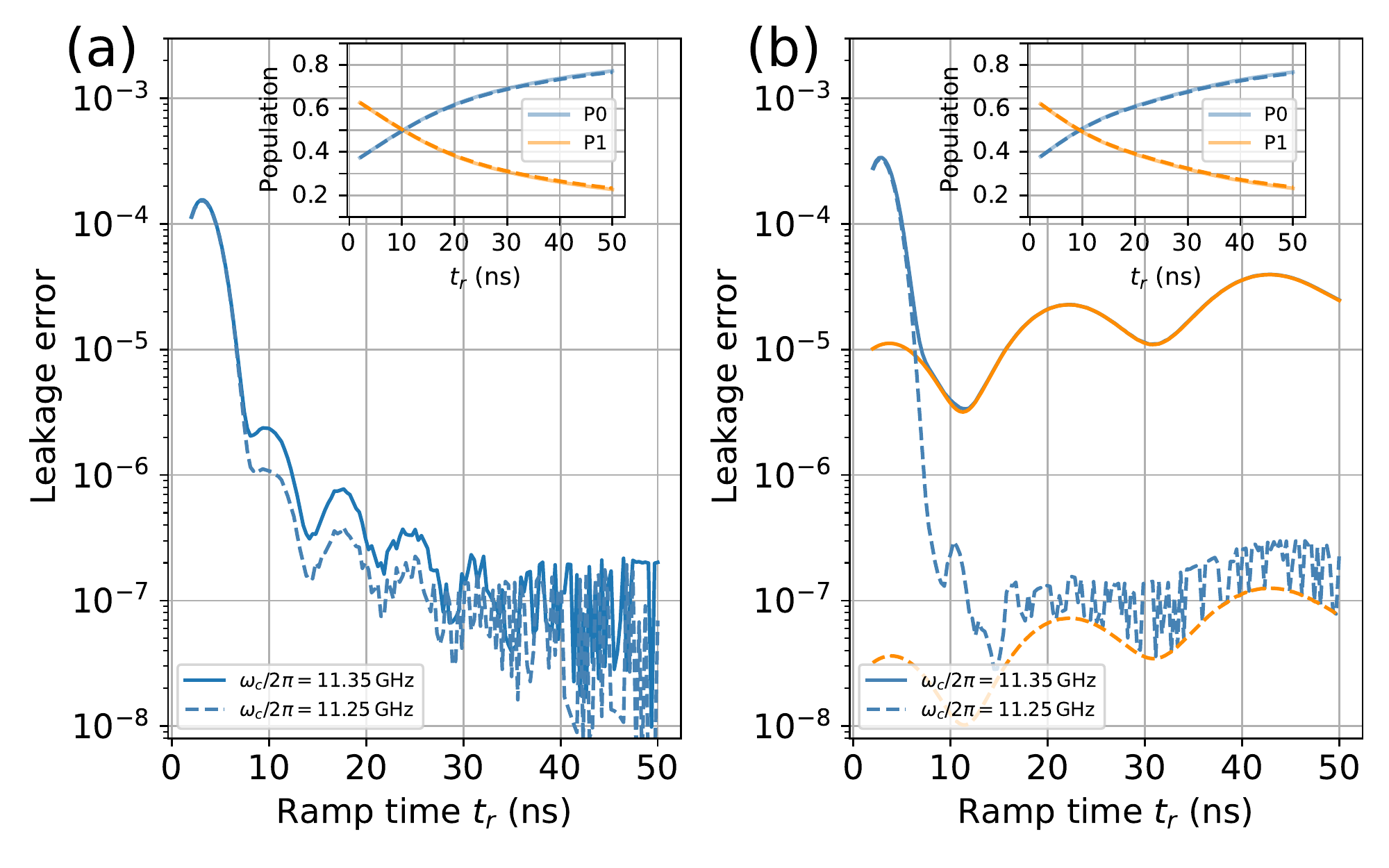}
\end{center}
\caption{Performing Z-controlled X rotations with the fast-adiabatic
flat-top pulse in the two-qubit system with tunable coupling. (a) Leakage as a function of the
pulse ramp time for qubit $Q_{0}$ initialized in state $|1\rangle$. Inset shows
the population in states $|0\rangle$ and $|1\rangle$ versus the ramp time for the
Z-controlled X rotations. The solid lines and dashed lines represent the results
with the coupler biased at two different idle points, i.e., $11.35\,\rm GHz$ and $11.25\,\rm GHz$, respectively.
Same as in Fig.~\ref{fig4}, here, the hold time of the utilized fast-adiabatic flat-top pulse is
fixed at $t_{h}=20\,\rm ns$. (b) Same as in (a), instead showing the results for $Q_{1}$.
Additionally, here also shows the population leakage to $Q_{0}$, as indicated by the orange lines.}
\label{fig8}
\end{figure}

Following the scheme introduced in Sec.~\ref{SecIIB}, here, Z-controlled single-qubit gates
are realized by using the fast-adiabatic flat-top pulse. Note here that in the present
two-qubit system, single-qubit gate operations for one qubit are tuned up and characterized
with the other qubit in its ground state $|0\rangle$.

Figure~\ref{fig8} shows the leakage versus the ramp time of
the pulse with the qubit initialized in its excited state $|1\rangle$. One can find that
while for $Q_{0}$, the result is in line with our theory discussed in
Sec.~\ref{SecIIB}, i.e., by increasing the ramp time, the leakage can be
further suppressed, the result of $Q_{1}$ seems unreasonable at first glance.
However, during gate operations applied to $Q_{1}$, $Q_{1}$ is
tuned from its idle point at $5.9\,\rm GHz$ to the working
point at about $6.1\,\rm GHz$, according to the fast-adiabatic pulse, while
$Q_{0}$ is fixed at its idle point at $6.0\,\rm GHz$. Therefore, during the pulse
ramp, $Q_{1}$ will sweep through a tiny avoided crossing formed by the
residual resonance XY coupling between $Q_{0}$ and $Q_{1}$ at $6.0\,\rm GHz$. On
contrast, during single-qubit gate operations, $Q_{0}$ will not sweep through $Q_{1}$.
As shown in Fig.~\ref{fig7}(a), the strength of the residual XY coupling
is about $0.1\,\rm MHz$. Thus, sweeping through this avoided
crossing slowly will generally cause more leakage into
the nearby qubit $Q_{1}$, as shown in Fig.~\ref{fig8}(b), where the
orange solid line denotes the population of $Q_{0}$ in state $|1\rangle$. One
can find that for $t_{r}\geq 10\,\rm ns$, the leakage into $Q_{0}$ gives
the leading contributions to the total leakage error. These results
suggest that there exists a trade-off
between gate error resulting from the qubit itself and error
from spectator qubits, i.e., suppressing leakage into $|2\rangle$
favors longer gate times, while mitigating the leakage into the $Q_{0}$
favors short gate times. This observation is in agreement with
that in previous work \cite{Zhao2022b}.

To address the above issue, one can change the idling coupler frequency, the XY coupling can thus be further
suppressed at the resonance point, i.e., $6.0\,\rm GHz$. By
biasing the coupler at $11.25\,\rm GHz$, the residual XY coupling is
suppressed below $0.01\,\rm MHz$. Accordingly, the leakage into $Q_{0}$ is indeed
suppressed heavily, as shown in Fig.~\ref{fig8}(b), where the dashed lines show
the results with the coupler biased at $11.25\,\rm GHz$.

Here, we turn to evaluate the gate performance of the baseband controlled
single-qubit gates, and use the metric of the state-average gate fidelity \cite{Pedersen2007}
in the following discussion (details on the fidelity calculation can also be found
in Ref.~\cite{Zhao2022b}). As mentioned in Sec.~\ref{SecIIA}, in the present work, we focus on the
implementation of $\sqrt{X}$ gates. From the inset of Figs.~\ref{fig8}(a)
and ~\ref{fig8}(b), one can find that for both qubits, $\sqrt{X}$ gates
can be realized with a ramp time of about $10\,\rm ns$, giving rise to the total gate
time of about $30\,\rm ns$. Moreover, even by biasing the coupler
at $11.35\,\rm GHz$, Figure~\ref{fig8} shows that when the ramp time
is about $10\,\rm ns$, the leakage error can still be suppressed
below $5\times10^{-5}$ for both qubits. This is to be
expected, since sweeping through the tiny avoided crossing with fast
speed could suppress leakage. By optimizing the ramp times, we find that
for both qubits, up to single-qubit Z rotations, $\sqrt{X}$ gates can be achieved
with gate fidelity exceeding $99.999\%$ (for $Q_{0}$, the gate fidelity is $99.9998\%$ and the
optimal gate time is $30.2\,\rm ns$, while for $Q_{1}$, are the $99.9996\%$
and $29.4\,\rm ns$). As mentioned before, Z gates can be easily realized by
choosing suitable time delays between flux pulses. In this way, universal
single-qubit gates can be achieved by combining Z gates and $\sqrt{X}$ gates.

\subsection{Two-qubit CZ gate}\label{SecIIIB}

\begin{figure}[tbp]
\begin{center}
\includegraphics[keepaspectratio=true,width=\columnwidth]{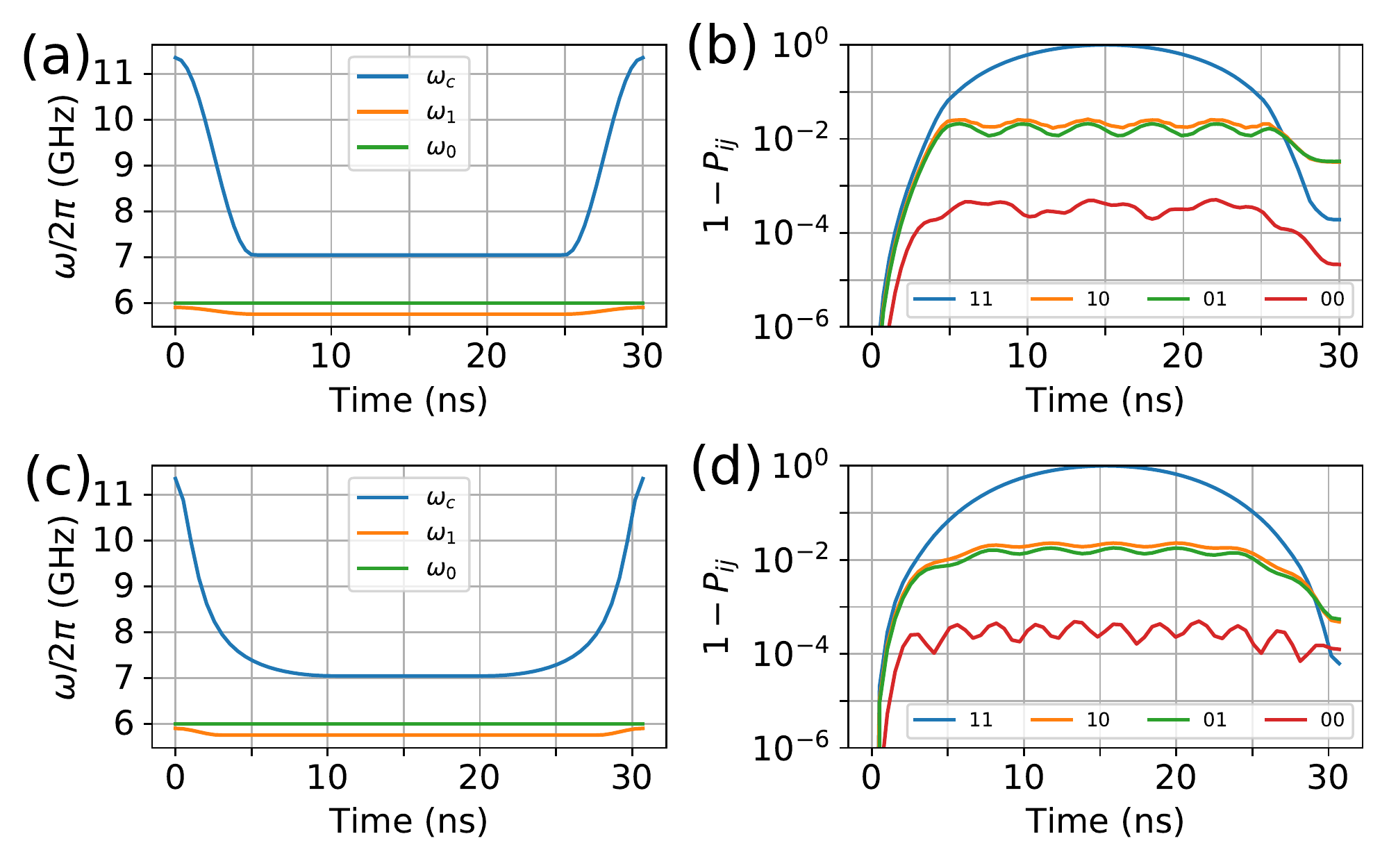}
\end{center}
\caption{Performing CZ gates with the fast-adiabatic flat-top pulse in the two-qubit
system with tunable coupling. During the gate operations, qubit $Q_{0}$ is fixed at its idle
point, i.e., $6.0\,\rm GHz$, qubit $Q_{1}$ and coupler $Q_{c}$ are tuned from their idle
points ($5.9\,\rm GHz$ and $11.35\,\rm GHz$) to the working points, resulting in a complete
population oscillation between states $|101\rangle$ and $|200\rangle$. (a) The cosine-decorated square
pulse with the ramp time of $10\,\rm ns$ for implementing CZ gates. Up to single-qubit phase
gates, the gate fidelity is $99.82\%$. (b)  Time evolution of the qubit
state population during the gate operation with the cosine-decorated square pulse.
Here, $P_{ij}$ denotes the population in state $|i0j\rangle$ for the two-qubit system initialized
in state $|i0j\rangle$. (c) For biasing the coupler, the fast-adiabatic flat-top pulse with a hold
time of $10\,\rm ns$ is employed, while for $Q_{1}$, a cosine-decorated square pulse with the ramp
time of $6\,\rm ns$ is used. Here, the total pulse length is $30.7\,\rm ns$ and up to single-qubit phase gates, the CZ gate
fidelity is $99.94\%$. (d) Time evolution of the qubit state population during the
gate operation with the fast-adiabatic pulse, showing that the population swap between
qubits is suppressed below $10^{-3}$.}
\label{fig9}
\end{figure}

Having discussed the single-qubit control, we now turn to the two-qubit case. Here, we
consider the implementation of CZ gates in the two-qubit system with an
always-on drive. During the gate operations, $Q_{0}$ is fixed at its idle
point, i.e., $6.0\,\rm GHz$, $Q_{1}$ is tuned from its idle
point ($5.9\,\rm GHz$) to the working point, where a complete oscillation between states $|101\rangle$
and $|200\rangle$ can occur. Meanwhile, the coupler is tuned from its idle
point at $11.35\,\rm GHz$ to a working point at about $7\,\rm GHz$, giving
rise to the CZ coupling strength of $20\,\rm MHz$ (see Appendix~\ref{D}).

Figure~\ref{fig9}(a) shows the typical control pulse, i.e., the pulse with a flat middle part
and cosine-shaped ramps (see Appendix~\ref{A} for details), with a pulse length of $30\rm ns$ for the CZ
implementation. By numerically optimizing the working points of $Q_{c}$
and $Q_{1}$, the gate fidelity of the implemented CZ gate (up to single-qubit Z phases) is $99.82\%$.
After inspecting the qubit dynamics during the gates, one can find that the leading error
source is the population swap between two qubits, as shown in Fig.~\ref{fig9}(b). Following
the fast-adiabatic scheme discussed in Sec.~\ref{SecIIB} and the previous work \cite{Martinis2014b,Sung2021}, here, the
fast-adiabatic flap-top pulses, as shown in Fig.~\ref{fig9}(c), with a
hold time of $10\,\rm ns$, is used to suppress the population swap. Accordingly, the
population swap is indeed largely suppressed, as shown in Fig.~\ref{fig9}(d), improving the
CZ gate fidelity to $99.94\%$ with a gate time of $30.7\,\rm ns$. Additionally, we note that
generally, by increasing the gate length, the residual gate error can be
further suppressed (see also in Appendix~\ref{E}).

The above results show that although there exists an always-on global drive, high-fidelity
two-qubit gates can still be achieved in a short time. This success is
mainly based on the fact that during the gate operations, the global drive is far detuned from both
qubits and the coupler.

\section{discussion}\label{SecIV}

Given the above theoretical analysis of the implementation of the baseband flux control in
tunable coupling architecture, in the following, we will give a few discussions of the challenges and opportunities for
realizing the baseband control strategy in large-scale superconducting quantum processors.

\subsection{Practical challenges}\label{SecIVA}

While our theoretical study shows that baseband controlled gate operations
can be realized with high fidelity and fast speed, we note that besides
the qubit decoherence, there exist several practical experimental issues
that will limit the available gate performance:

(i) Flux pulse distortion. Flux pulse distortion has been demonstrated as a critical issue faced by baseband
flux-controlled gate operations \cite{Jerger2019,Rol2020,Foxen2019}. Moreover, the above-demonstrated
high-fidelity gate operations are achieved by using pulse shaping technologies, thus, the
impact of flux pulse distortion can become more prominent in our setting.

(ii) Stray coupling beyond nearest neighbors. Generally, in our setting, the always-on drive is shared
by multiple qubits. When performing single-qubit gates in parallel, multi-qubit will
be tuned on-resonance with the same drive. This means that any stray coupling between these qubits
will cause population swaps among these qubits, as discussed in Sec.~\ref{SecIIIA}, leading to additional
gate errors compared to isolated gates. While near-neighbor couplings between qubits can be controlled well
in the tunable coupling architecture, parasitic coupling beyond nearest neighbors can still exist due to, such as
stray capacitive coupling, in multi-qubit systems \cite{Barends2014,Zajac2021,Yanay2022,Zhao2022b}. This will
degrade the efficiency of the baseband control strategy in large-scale quantum processors.

(iii) Defect modes, such as TLSs \cite{Muller2019}. Same as in (ii), when performing single-qubit gates, the
working frequencies of multi-qubit are almost limited to a fixed one, i.e., the frequency of the
shared drive. This will limit the ability to mitigate the impacts from defect modes by tuning the qubit away
from the defects \cite{Klimov2018}.

(iv) Keeping track of the single-qubit phase accumulation. In our setting, the qubit
frequency at its idle point is detuned from the always-on drive. Thus, the single-qubit
phase will accumulate at the speed of the drive detuning $\Delta$ during the idle time.
While the accumulated phase can be employed to realize single-qubit Z gates,
on the other hand, when performing gate sequences or quantum circuits, the accumulated phase
should be tracked carefully over the whole time domain. Compared with the traditional microwave
control, this could complicate the implementation of quantum circuits.

In addition, we note that owing to the great flexibility of Z-controlled $\sqrt{X}$ gates, as discussed
in Sec.\ref{SecIIA}, the issues, related to (ii) and (iii), may be addressed. From the results shown in
Fig.~\ref{fig2}, we can find that given a fixed drive amplitude or a fixed pulse
length, $\sqrt{X}$ gates can be achieved with a small drive detuning, for which its magnitude can
even be compared with that of the always-on drive. Thus, when implementing isolated or paralleled
single-qubit gates, the working frequencies of qubits can be biased intentionally at
different frequency points, thus impacts of sub-MHz stray coupling can be mitigated. Similarly,
the defect's impact can be suppressed by biasing qubits away from the leading
defect modes.

\begin{figure}[tbp]
\begin{center}
\includegraphics[keepaspectratio=true,width=\columnwidth]{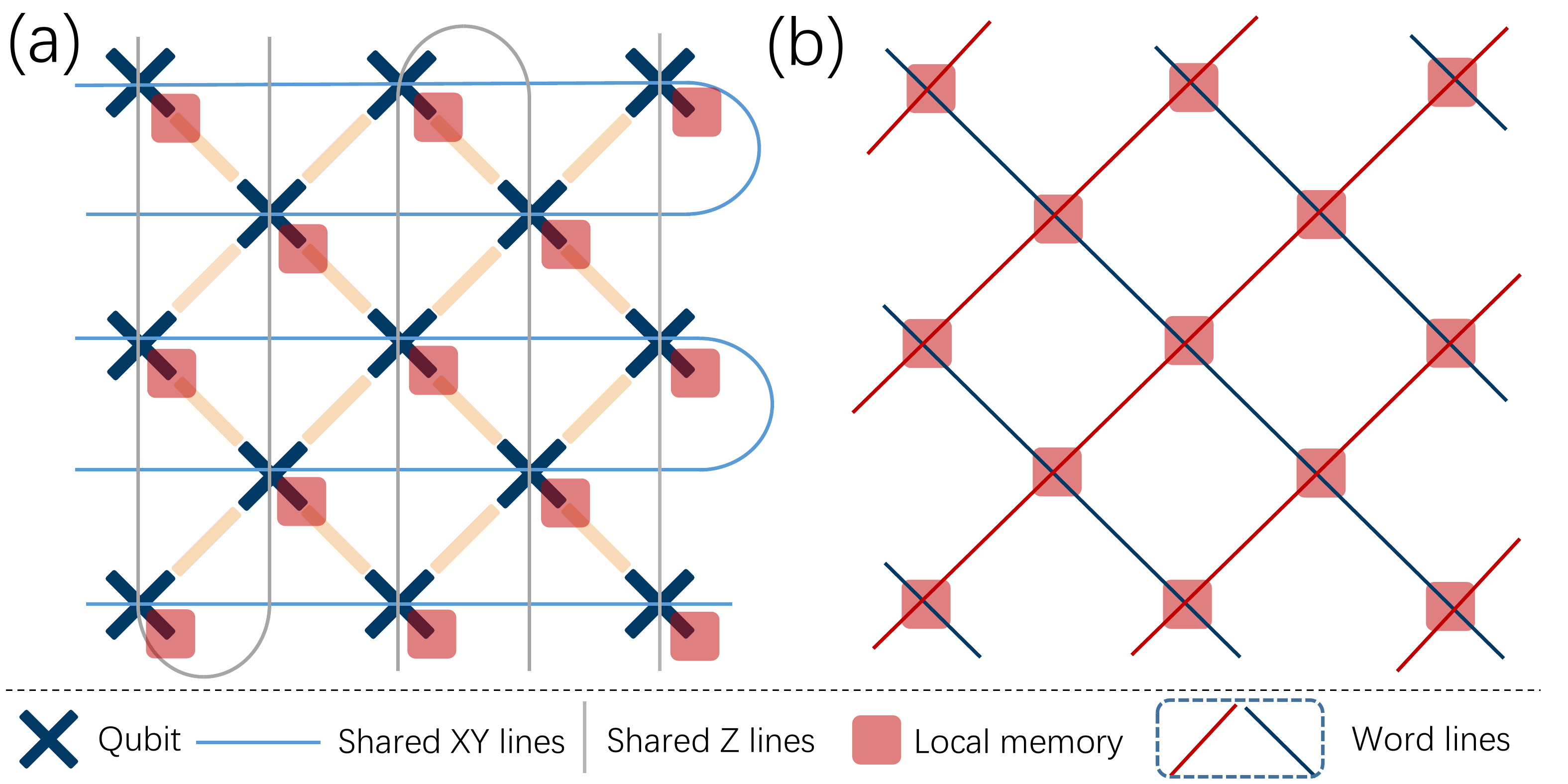}
\end{center}
\caption{(a) Multiplexing control of qubit lattices of frequency-tunable superconducting
qubits with shared XY and Z lines and local programmable memory. During the parallel gate operations, the
local memory can be used to switch on or off the control on the individual qubit, and can provide the
static bias for compensating qubit non-uniformity. (b) Network of word lines for digital addressing the
local memory.}
\label{fig10}
\end{figure}

\subsection{Opportunities for solving challenges towards large-scale quantum processors}\label{SecIVB}

Currently, in small-scale superconducting quantum processors, each qubit has its
dedicated control lines, such as XY lines and flux (Z) lines. Moreover, due to the
non-uniformity of qubit parameters, such as qubit frequency and anharmonicity, the
coupling efficiency between qubits and control lines, and the signal attenuation and
the distortion in control lines, control pulses can differ from qubit to qubit. Thus,
generally, microwave control pulses differ with each other in their amplitudes, frequencies,
and phases, while for the baseband flux pulse, their amplitudes could be different. When
scaling up to large-scale quantum computing, such strategy is not scalable.

Given the recent progress in the pursuit of scalable spin-based quantum
computing with multiplexing technologies and crossbar
technologies \cite{Hill2015,Vandersypen2017,Veldhorst2017,Li2018}, we may also
consider how to utilize these technologies for solving the above-mentioned
challenges toward large-scale superconducting quantum processors. One possible
example is schematically illustrated in Fig.~\ref{fig10}(a), where both the XY and Z lines
are shared by multiple qubits in a square lattice of frequency-tunable qubits.
Similarly, in qubit architectures with tunable couplers, the Z-line shared scheme could be
employed for controlling tunable couplers, thus enabling the implementation of
baseband-controlled two-qubit gates in parallel.

Note that in principle, the shared control
scheme should be applicable for driving all qubits with a single continuous microwave
source. As a practical application, we expect that the shared scheme could be feasible
for tens of superconducting qubits. With this strategy, the number of needed microwave drive lines
is, at least, an order of magnitude less than that in the traditional setting. Additionally, to
integrate with the widely used flip-chip technology, the shared drive may be applied
to multiple qubits through a shared XY line. According to the discussion given in
Ref.~\cite{Krinner2019}, we estimate that when the strength of the always-on microwave drive
is about 10 MHz ($\sqrt{X}$ gates with a gate time of 30 ns), the required power is
about -14 dBm at the room temperature, and through a series of attenuators and
filters (giving rise to the total attenuation of 60 dB), the actual power delivered to
the qubit chip is about -74 dBm. As in the traditional setting, the average required
power per qubit drive line is about -78 dBm (to the qubit chip)~\cite{Krinner2019}, we
expect that the heating issue of the present shared control scheme can be addressed.

Compared with the spin qubit, it seems that the superconducting qubit can provide
more flexible control over its physical size and qubit
parameters \cite{Martinis2020,Barends2014,Zhao2020N,Mamin2022,Zhao2022c,Chow2015}, yet, it
can also show prominent non-uniformity. Unfortunately, the success of the multiplexing
technologies and crossbar technologies highly hinges on the uniformity of qubit
parameters. This can be more prominent for superconducting quantum processors based
on individual microwave control. In the context of the implementation of multiplex
control of superconducting qubits, baseband flux control may alleviate this
issue of non-uniformity. Within our baseband control
setup, and applying the multiplexing technologies shown in Fig.~\ref{fig10}(a), there are three
main leading challenges from the qubit non-uniformity:

(i) The non-uniform amplitude of the shared microwave
drive or (ii) flux pulse felt by qubits (caused by, such as the different coupling
efficiency between qubits and the global XY/Z line and the signal attenuation in control
lines);

(iii) Independently calibrated parameters of flux pulse, including pulse length and pulse
shape, for implementing accurate control on individual qubits.

\begin{figure}[tbp]
\begin{center}
\includegraphics[keepaspectratio=true,width=\columnwidth]{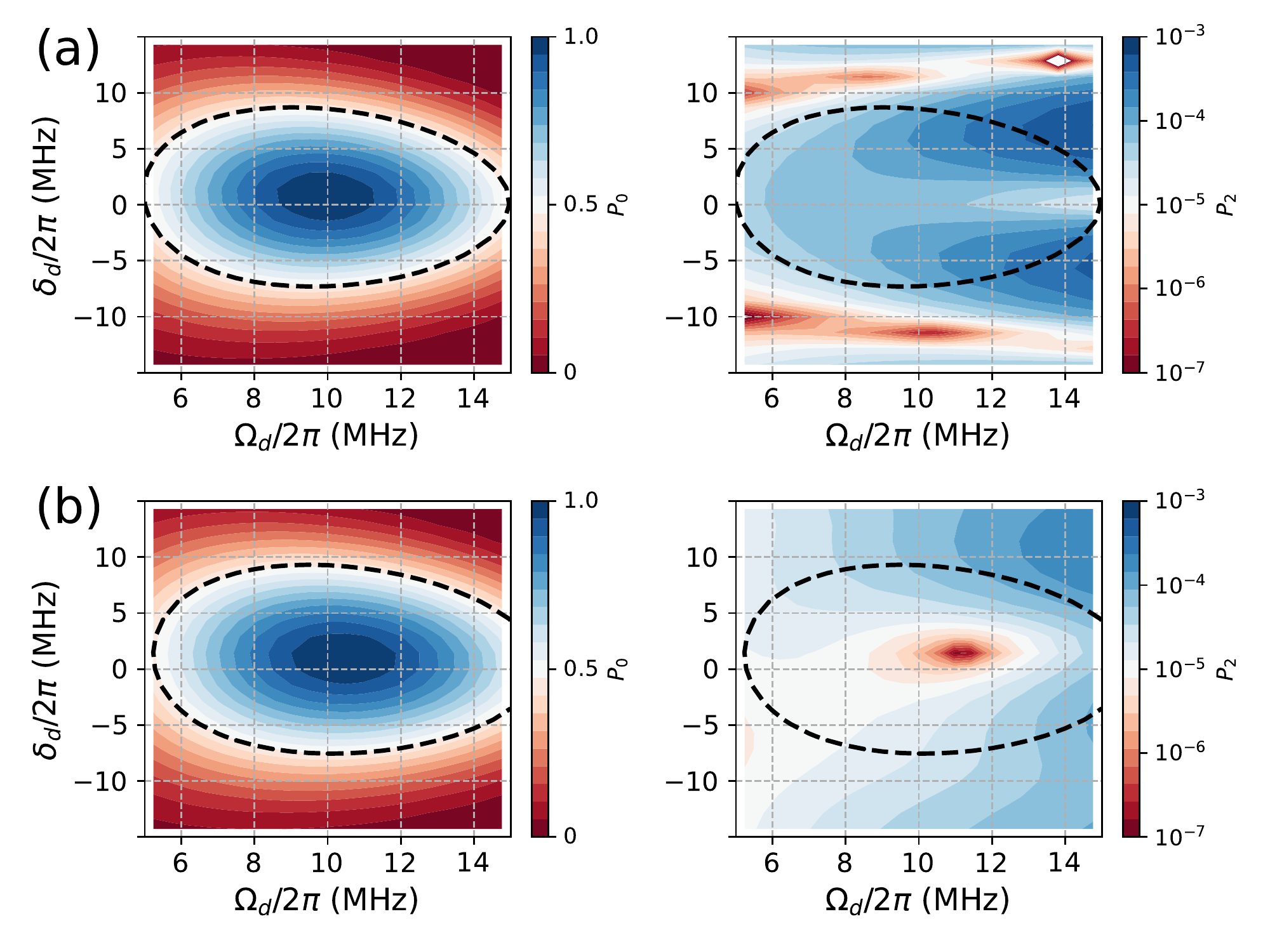}
\end{center}
\caption{Flexibility of $\sqrt{X}$ rotations on superconducting transmon qubits.
(a) Population in states $|0\rangle$ (left panel) and $|2\rangle$ (right panel) versus the
drive detuning and the drive amplitude with the qubit prepared in state $|1\rangle$.
Here, the qubit anharmonicity is $-250\,\rm MHz$, the drive detuning at the idle point is $-100\,\rm MHz$,
and the length of the square pulse is $50\,\rm ns$. The dashed lines indicate the
available parameter sets for implementing $\sqrt{X}$ rotations. (b) same as in (a),
instead showing the case with cosine-decorated square pulses. The ramp time and the hold
time are $10\,\rm ns$ and $50\,\rm ns$, respectively.}
\label{fig11}
\end{figure}

However, as discussed in Sec.\ref{SecIIA}, for two-level systems, given a fixed pulse
length and pulse shape (i.e., square shape, see Appendix~\ref{A} for results with smooth
pulses), owing to the flexibility of Z-controlled single-qubit gates (based on $\sqrt{X}$ gates),
the available parameter ranges (i.e., the drive amplitude and the drive detuning) can
be explored for compensating the non-uniform of drive amplitude. This exciting feature
is illustrated in Fig.~\ref{fig2}(b). Furthermore, similar results can also be obtained
for superconducting qubits, such as transmon qubits. Figures~\ref{fig11}(a) and~\ref{fig11}(b)
present the results for transmon qubits with $50$-$\rm ns$ square pulses and smooth pulses (i.e., cosine-decorated
square pulse with a hold time of $50\,\rm ns$ and a ramp time of $10\,\rm ns$), respectively.
Here, the qubit anharmonicity is $-250\,\rm MHz$ and the other used parameters are same as
in Fig.~\ref{fig2}. In addition, we note that to achieve uniform control pulses, the Z gate scheme based on
time-delay and the proposed fast-adiabatic scheme cannot be employed. Here, we can instead use
cosine-decorated square pulses for implementing Z gates by tuning the qubit from the idle
point, and for suppressing leakage. From Figs.~\ref{fig11}(a) and~\ref{fig11}(b), one can find that
by adding cosine-shaped ramps, the leakage error can be suppressed below $10^{-4}$, while for
the square pulse the leakage can approach $10^{-3}$. To further suppress the leakage
error, one can increase the ramp time or decrease the drive amplitude. However, this will
increase gate length and thus cause more decoherence errors.

Considering the above results, the above three challenges can be effectively overcome by
only addressing the non-uniformity issue of flux pulse amplitudes. This non-uniformity issue
could be removed by developing on-chip programmable memory, such as the one demonstrated by using
Single Flux Quantum (SFQ) logic \cite{Johnson2010,McDermott2010}, which could be used to compensate for the
remaining non-uniformity. In this way, combined with the word lines for digital
addressing, as shown in Fig.~\ref{fig10}(b), it is possible using only a few global XY and Z lines
to achieve parallel control of large numbers
of qubits \cite{Hill2015,Vandersypen2017,Veldhorst2017,Li2018}. Nevertheless,
given the practical experimental limitations, such as the limited cooling
power, the realization of the local programmable memory, which is compatible
with superconducting qubits, is still rarely explored \cite{Johnson2010}, and
undoubtedly, will be one of the most crucial challenges for implementing multiplexing
control technologies. Last, but not least, we must stress
that before solving the issue of pulse distortion, the efficiency of the above-discussed
scheme could be limited for implementing gate-based quantum computing.

\section{conclusion}\label{SecV}

In this work, we propose and theoretically study the possibility of implementing baseband
control of superconducting qubits, which are subjected to an always-on global drive. Our
results provide a general understanding and the basic principles of realizing
the baseband control scheme for superconducting qubits, such as frequency-tunable transmon
qubits. In the qubit architecture with tunable coupling, we show that high-fidelity and
fast-speed gate operations are possible by employing this baseband control scheme. Additionally,
we have further discussed potential challenges and opportunities for implementing such baseband
control strategy toward large-scale superconducting quantum processors.

\begin{acknowledgments}
We acknowledge helpful discussions with Zhaohua Yang, Yanwu Gu, and Zhi-Hai Liu. This work
was supported by the National Natural Science Foundation of
China (Grants No.12204050, No.11890704, and No.11905100), the Beijing
Natural Science Foundation (Grant No.Z190012), and the
Key-Area Research and Development Program of Guang Dong Province (Grant No. 2018B030326001).
P.X. was supported by the National Natural Science Foundation of
China (Grant Nos. 12105146, 12175104).
\end{acknowledgments}

\emph{Note added.}-- During the preparation of this manuscript, we became aware of a recent related work \cite{Bejanin2022},
which presents the experimental demonstration of baseband-controlled single-qubit gates in superconducting transmon qubits.

\appendix

\section{Flexibility of $\sqrt{X}$ rotations}\label{A}

\begin{figure}[tbp]
\begin{center}
\includegraphics[keepaspectratio=true,width=\columnwidth]{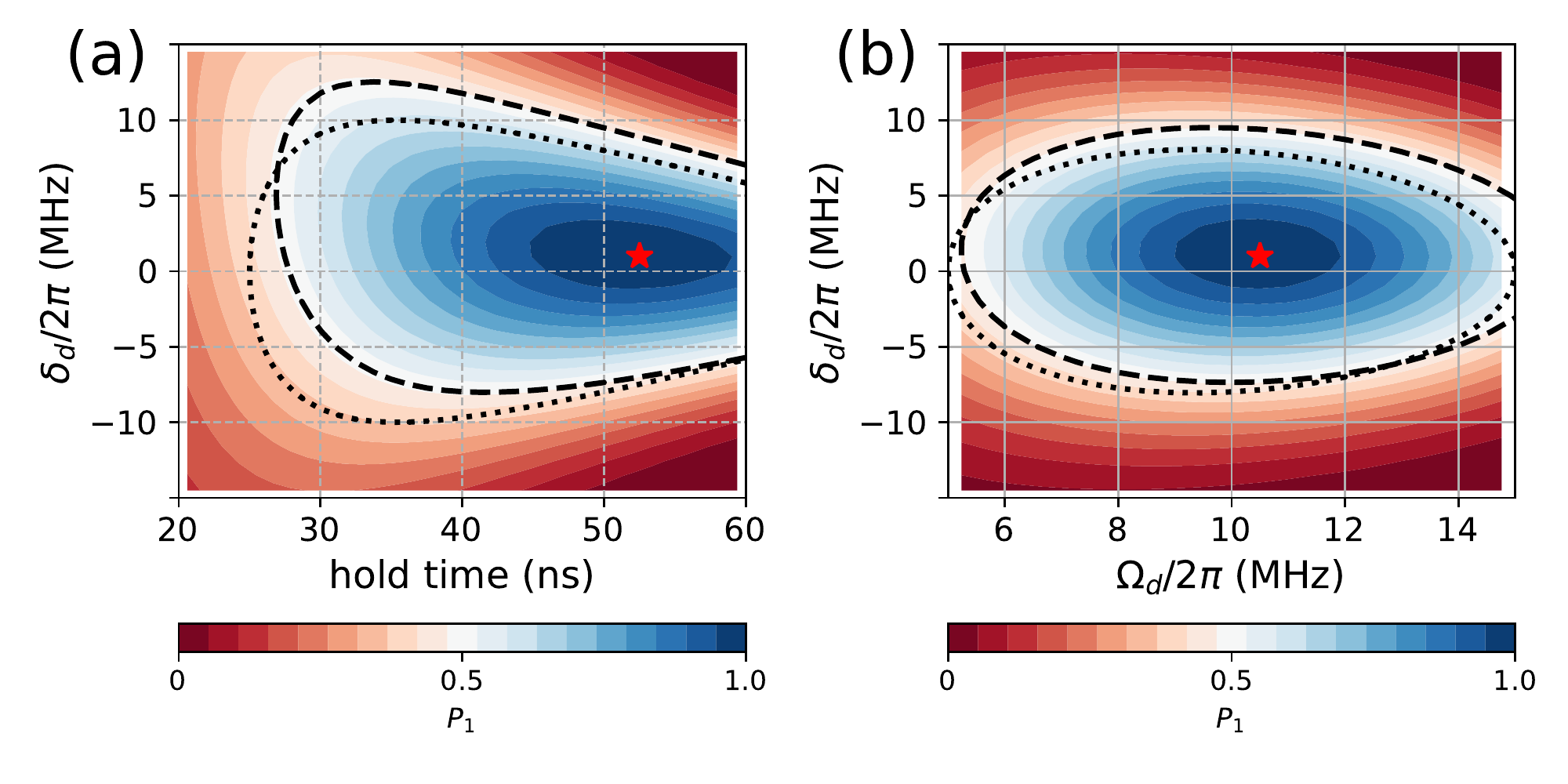}
\end{center}
\caption{Same as in Fig.~\ref{fig2}, instead showing results with cosine-decorated square pulses.
Here, the hold time is the pulse length of the flat part of the cosine-decorated square pulse, and the
ramp time of the pulse is fixed at $10\,\rm ns$. }
\label{fig12}
\end{figure}

\begin{figure}[tbp]
\begin{center}
\includegraphics[keepaspectratio=true,width=\columnwidth]{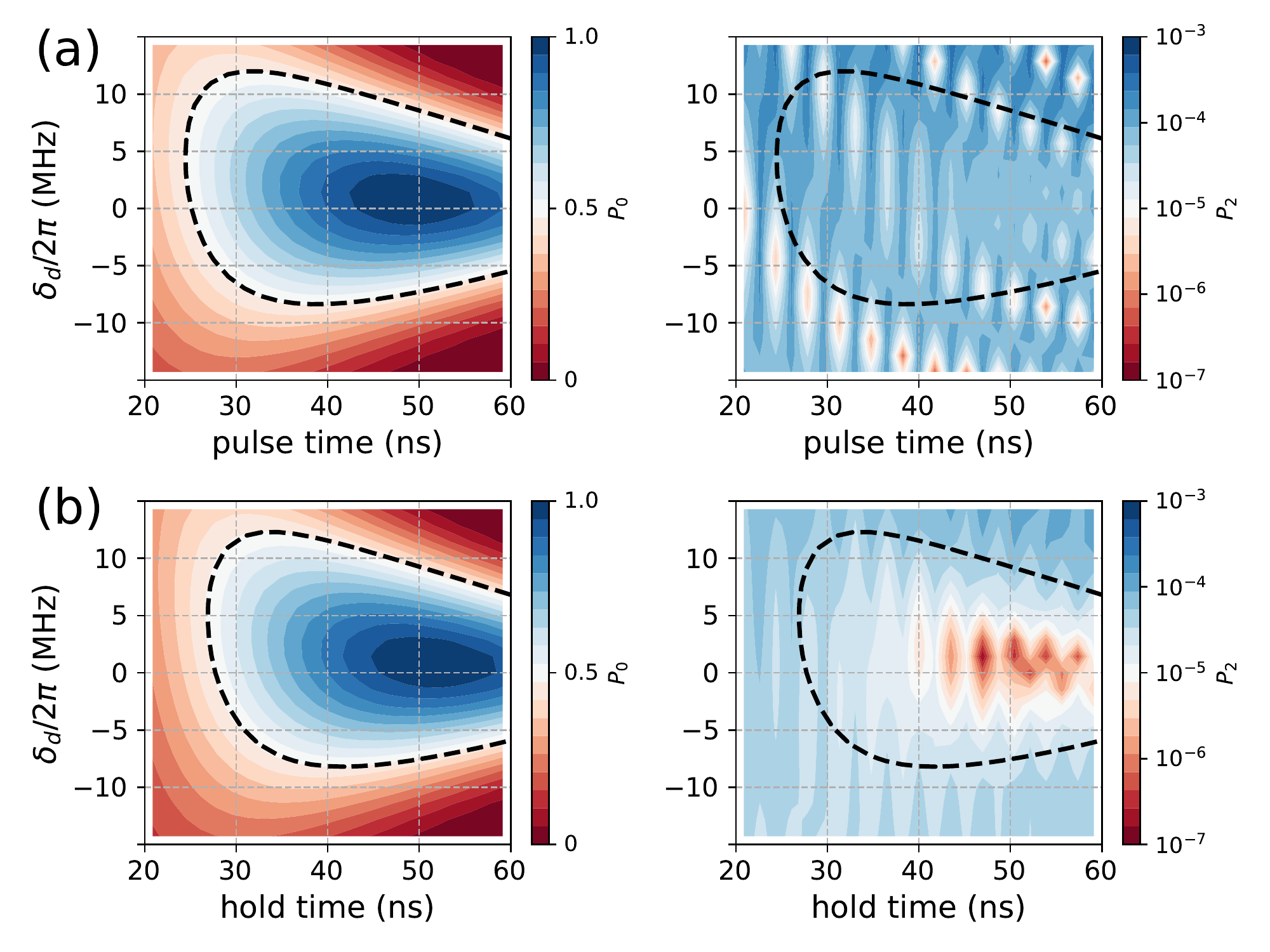}
\end{center}
\caption{Same as in Figs.~\ref{fig2}(a) and~\ref{fig12}(a), instead showing the results for the transmon qubit
with the anharmonicity of $-250\,\rm MHz$. (a) Population in states $|0\rangle$ (left panel)
and $|2\rangle$ (right panel) versus the drive detuning and the pulse time with the qubit
prepared in state $|1\rangle$. Here, the drive detuning at the idle point is $-100\,\rm MHz$,
and the drive amplitude is $10\,\rm MHz$. The dashed lines indicate the
available parameter sets for implementing $\sqrt{X}$ rotations. (b) same as in (a),
instead showing the case with cosine-decorated square pulses. The ramp time is $10\,\rm ns$.}
\label{fig13}
\end{figure}

In Fig.~\ref{fig2}, we show the dynamics of Z-controlled two-level systems subjected to
an always-on drive. Here, as shown in Fig.~\ref{fig12}, we further present the result for the case
with cosine-decorated square pulses, i.e.,
\begin{align}
\Delta(t)\equiv
\begin{cases}
\Delta_Z[1-\cos{(2\pi \frac{t}{t_r}})]/2  \;, &0<t<t_r/2\\
\Delta_Z\;,  &t_r/2<t<t_g-t_r/2\\
\Delta_Z[1-\cos{(2\pi \frac{t_g-t}{t_r}})]/2 \;, &t_g-t_r/2<t<t_g
\end{cases}
\label{eqa1}
\end{align}
where, $\Delta_Z$ denotes the peak pulse amplitude, $t_r$ is the ramp time, and $t_g$ represents
the total pulse length. One can find that same as that with square pulses, to implement X rotations, a
small overshoot is needed, as indicated by the red stars. Accordingly, as shown in Fig.~\ref{fig13},
we also give the results, including both the population in state $|0\rangle$ and the leakage
into state $|2\rangle$, for the transmon qubit. Here the qubit anharmonicity is $-250\,\rm MHz$, and
the transmon qubit is prepared in state $|1\rangle$.

\section{fast-adiabatic pulse}\label{B}

As discussed in the main text, the fast-adiabatic scheme is employed for designing
an optimal control pulse for suppressing
leakage \cite{Martinis2014b}. Our strategy is using the Slepian-based method to design an
optimal ramp pulse, and then inserting a square pulse into the optimal pulse. In
this way, the target fast-adiabatic flat-top pulse is synthesized for implementing
low-leakage X rotations. Therefore, here we only focus on finding the optimal
ramp pulse. As shown in Fig.~\ref{fig3}(a), the leakage error results from the non-adiabatic
evolution in the leakage subspace. Generally, during X rotations i.e., the drive detuning
varies from the idle point with the control angle $\theta_{i}=\arctan{[\sqrt{2}\Omega_{d}/(\Delta_{d}+\alpha_{q})]}$
to the working point with the control angle $\theta_{f}=\arctan{[\sqrt{2}\Omega_{d}/\alpha_{q}]}$, and
then comes back. Following Ref.~\cite{Martinis2014b}, the ramp pulse with a length of $t_{g}$ can be
parameterized in terms of Fourier basis functions, and is given by

\begin{eqnarray}
\begin{aligned}\label{eqb1}
\theta(t)=\theta_{i}+ \frac{\theta_{f}-\theta_{i}}{2}\sum\limits_{n=1,2,3...}\lambda_{n}\left[1-\cos\frac{2n\pi t}{t_{g}}\right]
\end{aligned}
\end{eqnarray}
with constraints on the coefficients $\Sigma_{n\,\rm{odd}}\,\,\lambda_{n}=1$. Here, to find the
optimal pulse, we consider keeping three Fourier terms. Following Ref.~\cite{Martinis2014b}, the
three optimized coefficients can be obtained numerically by minimizing the integrated spectral
density of the pulse above a chosen frequency.

\begin{figure}[tbp]
\begin{center}
\includegraphics[keepaspectratio=true,width=\columnwidth]{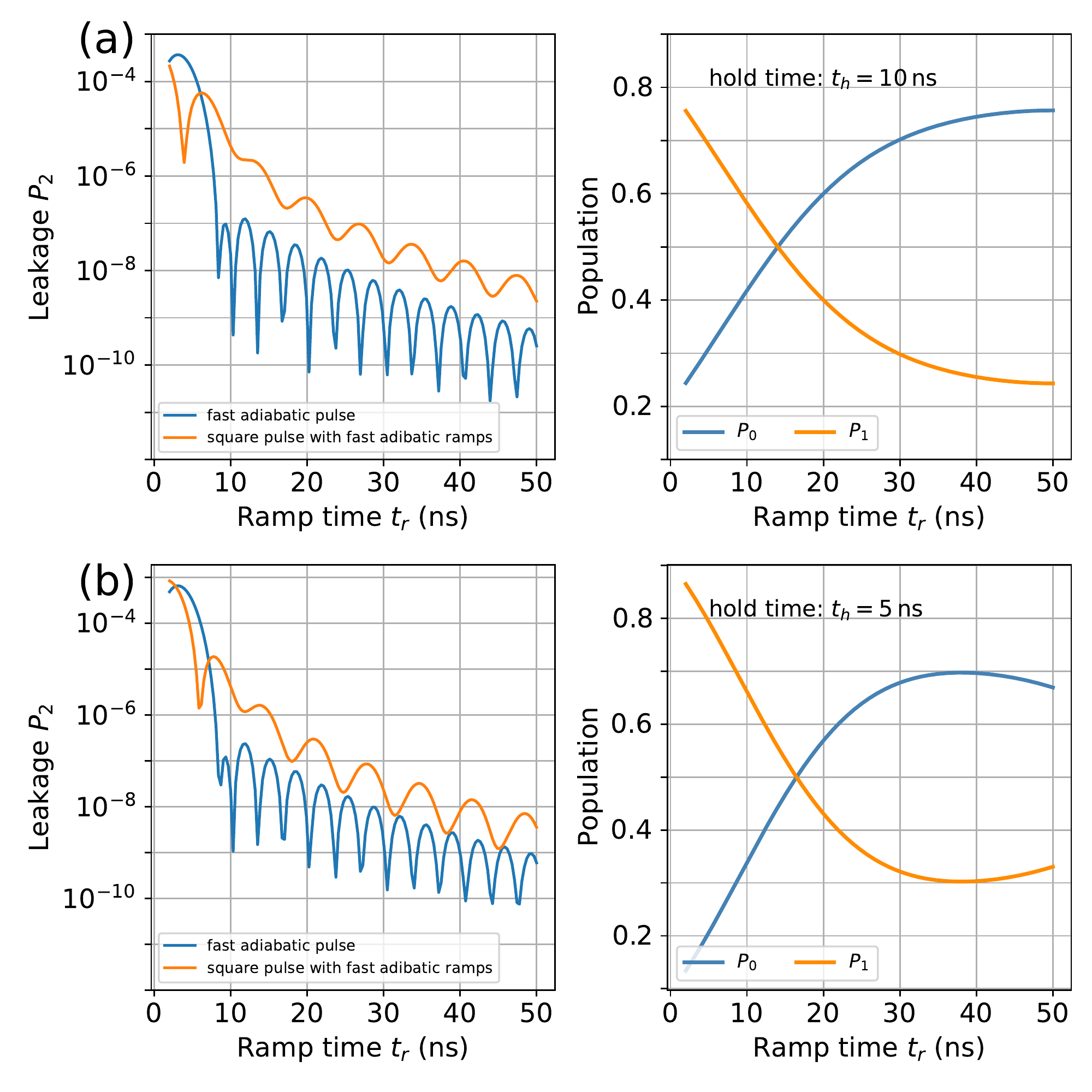}
\end{center}
\caption{leakage with varying driving strengths. Same as in Fig.~\ref{fig4}, here shows
the qubit population and leakage versus the ramp time for the transmon qubit initialized in
state $|1\rangle$ with the anharmonicity $\alpha_{q}/2\pi=-250\,\rm MHz$. (a) The drive
amplitude is $\Omega_{d}/2\pi=15\,\rm MHz$ and the hold time is fixed at $10\,\rm ns$.
(b) The drive amplitude is $\Omega_{d}/2\pi=20\,\rm MHz$ and the hold time is
fixed at $5\,\rm ns$. The other system parameters are the same as in Fig.~\ref{fig4}.}
\label{fig14}
\end{figure}

In the main text, the efficiency of the above-discussed optimal pulse is only evaluated with
the drive amplitude of $10\,\rm MHz$. Here, we present more results with different
drive strengths. In Figs.~\ref{fig14}(a) and~\ref{fig14}(b), we show the results
with $\Omega_{d}/2\pi=15\,\rm MHz$ and $\Omega_{d}/2\pi=20\,\rm MHz$, respectively. Similar to the
results shown in Fig.~\ref{fig4}, we find that fast-speed $\sqrt{X}$ rotations can be realized
with low leakage.

\section{readout}\label{C}

Here for easy reference, following Ref.~\cite{Walter2017}, we briefly describe
the processing of the recording obtained from continuous measurement to infer qubit states.
As mentioned in the main text, by encoding the qubit information into
single quadrature, i.e., in-phase quadrature $I$, the qubit state can be
inferred by recoding $I_{t}=\langle a_{r}^{\dagger}+a_{r}\rangle(t)$ during
the continuous measurement. To maximize the readout
fidelity with a given measurement time (here is $t_i=250\,\rm ns$), the records
of $I_{t}$ are integrated over the measurement time $t_{i}$ with a weight function, giving
rise to the integrated readout quadrature value

\begin{equation}
\begin{aligned}\label{eqc1}
I=\sqrt{k}\int_{0}^{t_i}W_{t}[I_t-\langle I_{t}^{(0)}\rangle]dt
\end{aligned}
\end{equation}
with the weight function
\begin{equation}
\begin{aligned}\label{eqc2}
W_{t}\varpropto|\langle I_{t}^{(1)}\rangle- \langle I_{t}^{(0)}\rangle|\, {\rm with\,} \int_{0}^{t_i}W_{t}^2 dt=1.
\end{aligned}
\end{equation}

Here, $\langle I_{t}^{(0)}\rangle$ and $\langle I_{t}^{(1)}\rangle$ represent the expectation values of $I_{t}$ for the qubit
prepared in $|0\rangle$ and $|1\rangle$, respectively.

\section{RWA and corrections}\label{D}

\begin{figure}[tbp]
\begin{center}
\includegraphics[keepaspectratio=true,width=\columnwidth]{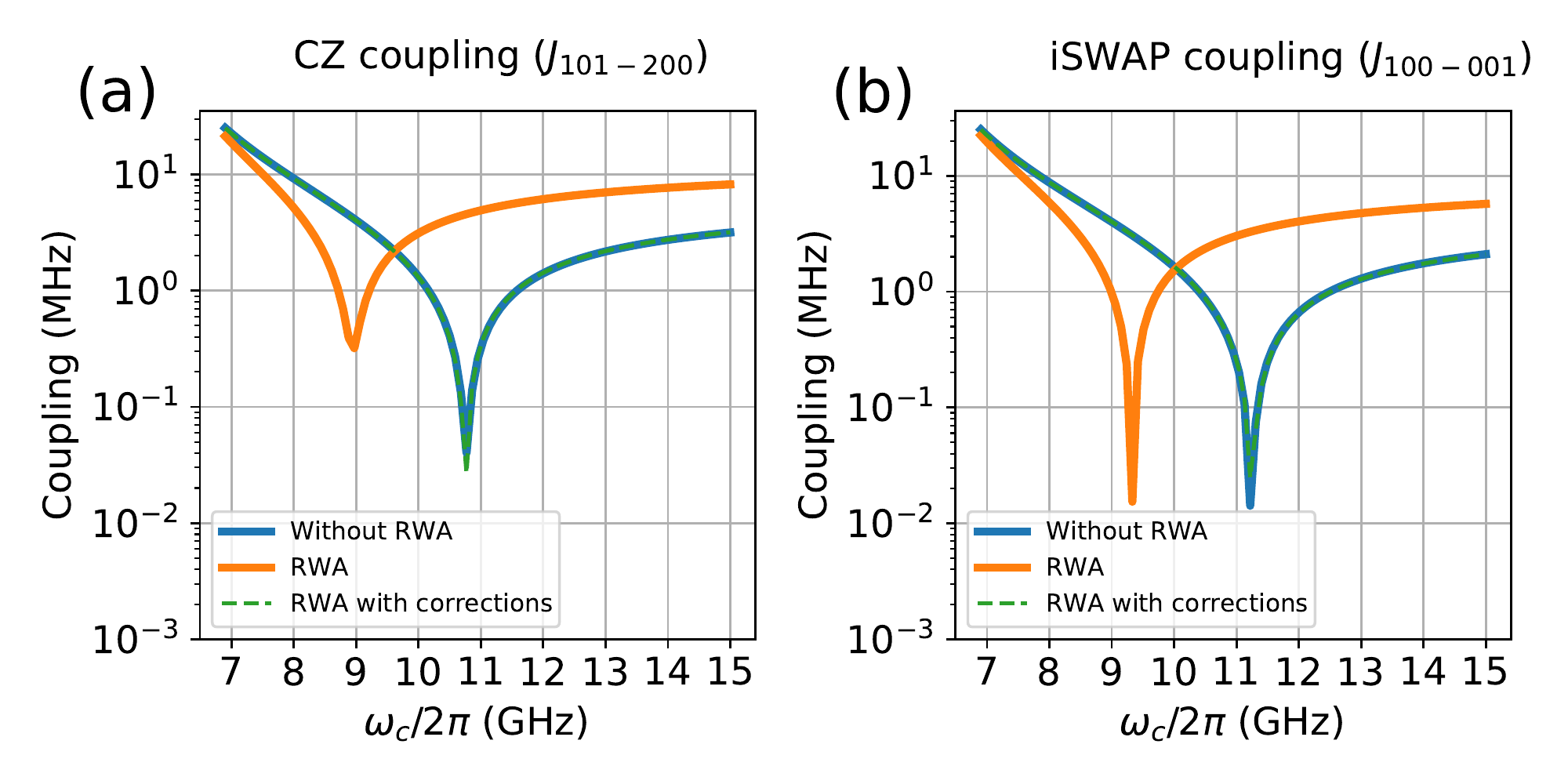}
\end{center}
\caption{Numerically calculated XY coupling strengths of (a) CZ coupling $J_{101-200}$ (the resonance coupling
between $|101\rangle$ and $|200\rangle$) and (b) iSWAP coupling $J_{100-001}$ (the resonance coupling
between $|100\rangle$ and $|001\rangle$). The blue and orange solid lines show the results without the RWA and the
results with the RWA, respectively, while the green dashed lines denote results with the RWA and the corrections.}
\label{fig15}
\end{figure}

For two frequency-tunable transmon qubits ($Q_{0}$ and $Q_{1}$) coupled via a tunable coupler $Q_{c}$, the system
Hamiltonian is given by
\begin{equation}
\begin{aligned}\label{eqd1}
H_{0}=&\sum_{j=0,1,c}\big(\omega_{j}a_{j}^{\dagger}a_{j}+\frac{\alpha_{j}}{2}a_{j}^{\dagger}a_{j}^{\dagger}a_{j}a_{j}\big)
\\&+\sum_{\substack{k=0,1,c\\j\neq k}}g_{jk}(a_{j}a_{k}^{\dagger}+a_{j}^{\dagger}a_{k})
\\&+\sum_{\substack{k=0,1,c\\j\neq k}}g_{jk}(a_{j}a_{k}+a_{j}^{\dagger}a_{k}^{\dagger}),
\end{aligned}
\end{equation}
After applying the RWA, the non-RWA term, i.e., the terms in the third line of Eq.~(\ref{eqd1}),
is omitted, giving rise to
\begin{equation}
\begin{aligned}\label{eqd2}
H_{\rm RWA}=&\sum_{j=0,1,c}\big(\omega_{j}a_{j}^{\dagger}a_{j}+\frac{\alpha_{j}}{2}a_{j}^{\dagger}a_{j}^{\dagger}a_{j}a_{j}\big)
\\&+\sum_{\substack{k=0,1,c\\j\neq k}}g_{jk}(a_{j}a_{k}^{\dagger}+a_{j}^{\dagger}a_{k})
\end{aligned}
\end{equation}

As mentioned in the main text, the non-RWA term can significantly affect the effective coupling
between qubits and shift the bare qubit frequency. This can be found in Fig.~\ref{fig15}, where
the numerically calculated XY coupling strengths for iSWAP coupling $J_{100-001}$ (the resonance coupling
between $|100\rangle$ and $|001\rangle$) and CZ coupling $J_{101-200}$ (the resonance coupling
between $|101\rangle$ and $|200\rangle$) are presented based on Eq.~(\ref{eqd1}) and Eq.~(\ref{eqd2}).

To remove the discrepancy within the RWA formalism, we consider eliminating the non-RWA terms
in Eq.~(\ref{eqd1}) by using the unitary transformation \cite{Zueco2009,Yan2018}
\begin{equation}
\begin{aligned}\label{eqd3}
U=\exp{\left[-\sum_{j\neq k}\frac{g_{jk}}{\Sigma_{jk}}(a_{j}a_{k}-a_{j}^{\dagger}a_{k}^{\dagger})\right]}
\end{aligned}
\end{equation}
with $\Sigma_{jk}=\omega_{j}+\omega_{k}$. This gives $H_{\rm RWA_C}=U^{\dag}H_{0}U$.
Expanding the above equation and keeping term up to second-order in the small parameters $g_{ic}/\Sigma_{i}$, we have
\begin{equation}
\begin{aligned}\label{eqd4}
H_{\rm RWA_C}\approx&\sum_{j=0,1,c}\big(\tilde{\omega}_{j}a_{j}^{\dagger}a_{j}+\frac{\alpha_{j}}{2}a_{j}^{\dagger}a_{j}^{\dagger}a_{j}a_{j}\big)
\\&+\sum_{\substack{i=0,1}}g_{ic}(a_{i}a_{c}^{\dagger}+a_{i}^{\dagger}a_{c})
\\&+\tilde{g}_{01}(a_{0}a_{1}^{\dagger}+a_{0}^{\dagger}a_{1})
\end{aligned}
\end{equation}
with the renormalized qubit frequency and coupler frequency
\begin{equation}
\begin{aligned}\label{eqd5}
&\tilde{\omega}_{i}=\omega_{i}-\frac{g_{01}^{2}}{\Sigma_{01}}-\frac{g_{ic}^{2}}{\Sigma_{ic}},\, {\rm with}\,i=0,1
\\&\tilde{\omega}_{c}=\omega_{c}-\frac{g_{0c}^{2}}{\Sigma_{0c}}-\frac{g_{1c}^{2}}{\Sigma_{1c}},
\end{aligned}
\end{equation}
and the effective strength of qubit-qubit coupling
\begin{equation}
\begin{aligned}\label{eqd6}
\tilde{g}_{01}=g_{01}-\frac{g_{0c}g_{1c}}{2}\left(\frac{1}{\Sigma_{0c}}+\frac{1}{\Sigma_{1c}}\right).
\end{aligned}
\end{equation}
Taking the above-obtained corrections into consideration, the results (dashed lines) with the
RWA show a great agreement with the results without the RWA, as shown in Fig.~\ref{fig15}.
Finally, we note that in our numerical analysis, each transmon qubit is modeled
as an anharmonic oscillator truncated with five levels. Additionally, to further reduce the computational
expenses, we further project the full system Hamiltonian to a smaller subspace where
at most five excitations are permitted. We justify this choice by checking the
resulting gate error compared to models with more energy levels or excitations
and find that the variation of the gate error is below $10^{-6}$.

\section{CZ gate error}\label{E}

\begin{figure}[tbp]
\begin{center}
\includegraphics[keepaspectratio=true,width=\columnwidth]{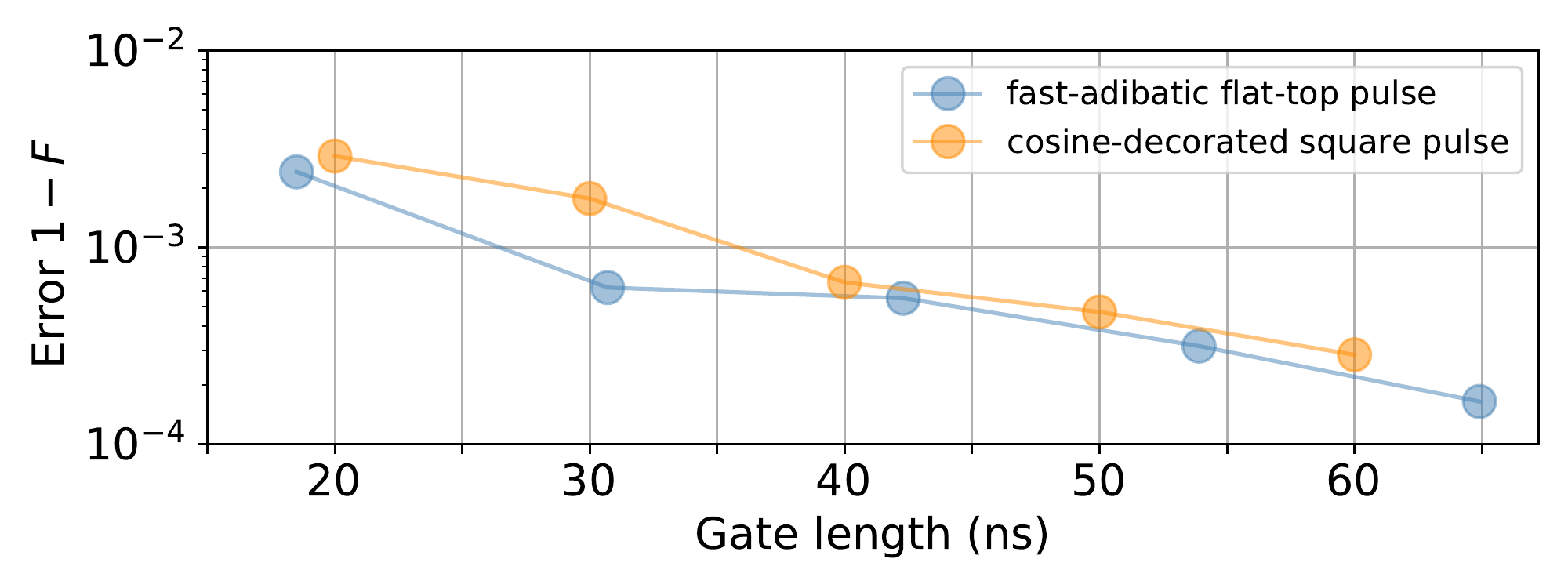}
\end{center}
\caption{Gate error versus the gate length. The blue and orange lines denote the
errors of CZ gates using the fast-adiabatic flat-top pulse and the cosine-decorated
square pulse. As in Fig.~\ref{fig9}(a), the ramp time of the cosine-decorated square
pulse is $10\,\rm ns$. The hold times of the fast-adiabatic flat-top pulses
are $\{0,\,10,\,20,\,30,\,40\}\,{\rm ns}$.}
\label{fig16}
\end{figure}

As mentioned in the main text, generally, increasing the gate length can further suppress the parasitic
interaction (i.e., qubit-qubit swap interactions and qubit-coupler swap interactions) induced gate
errors. Here, we provide more numerical results on this issue. Figure~\ref{fig16} shows the gate error
versus gate length for the discussed two types of pulses, i.e., the fast-adiabatic flat-top pulse and
the cosine-decorated square pulse.

\end{document}